\RequirePackage{ifpdf}
\documentclass[letterpaper]{JHEP3}
\usepackage{epsfig}
\usepackage{comment}
\usepackage{amsmath}

\def\ba{\begin{eqnarray}}
\def\ea{\end{eqnarray}}
\def\be{\begin{equation}}
\def\ee{\end{equation}}
\def\nn{\nonumber}
\def\exd{{\rm d}}
\def\pd{\partial}
\makeatletter
\def\x@arrow{\DOTSB\Relbar}
\def\xlongequalsignfill@{\arrowfill@\x@arrow\Relbar\x@arrow}
\newcommand{\xlongequal}[2]{%
    \ext@arrow 0099\xlongequalsignfill@{#1}{#2}}
\makeatother

\newcommand{\roughly}[1]{\mathrel{\raise.3ex\hbox{$#1$\kern-0.85em
\lower1ex\hbox{$\sim$}}}}

\newcommand{\lsim}{\roughly<}
\newcommand{\gsim}{\roughly>}

\def\nott#1{\setbox0=\hbox{$#1$}                % set a box for #1
   \dimen0=\wd0                                 % and get its size
   \setbox1=\hbox{/} \dimen1=\wd1               % get size of /
   \ifdim\dimen0>\dimen1                        % #1 is bigger
      \rlap{\hbox to \dimen0{\hfil/\hfil}}      % so center / in box
      #1                                        % and print #1
   \else                                        % / is bigger
      \rlap{\hbox to \dimen1{\hfil$#1$\hfil}}   % so center #1
      /                                         % and print /
   \fi}                                         %

\def\endignore{}
\def\ignore #1\endignore{} % use to "comment out" text

\def\be{\begin{equation}}
\def\beq\begin{equation}
\def\ee{\end{equation}}
\def\bea{\begin{eqnarray}}
\def\eea{\end{eqnarray}}

\def\eqa{\begin{eqnarray}}
\def\eeqa{\end{eqnarray}}
\def\eq{\begin{equation}}
\def\eeq{\end{equation}}

\newbox\charbox
\newbox\slabox
\def\slsh#1{{      % Feynman slash
        \setbox\charbox=\hbox{$#1$}
        \setbox\slabox=\hbox{$/$}
        \dimen\charbox=\ht\slabox
        \advance\dimen\charbox by -\dp\slabox
        \advance\dimen\charbox by -\ht\charbox
        \advance\dimen\charbox by \dp\charbox
        \divide\dimen\charbox by 2
        \raise-\dimen\charbox\hbox to \wd\charbox{\hss/\hss}
        \llap{$#1$}
}}

\def\nn{\nonumber}

\def\pref#1{(\ref{#1})}

\def\ol#1{\overline{#1}}

\def\exd{{\rm d}}
\def\nn{\nonumber}
\def\pref#1{(\ref{#1})}
\def\be{\begin{equation}}
\def\ee{\end{equation}}

\def\beq{\begin{equation}}
\def\eeq{\end{equation}}
\def\beqa{\begin{eqnarray}}
\def\eeqa{\end{eqnarray}}

\def\cC{{\cal C}}

\def\cE{{\cal E}}
\def\cF{{\cal F}}
\def\cG{{\cal G}}
\def\cH{{\cal H}}

\def\cJ{{\cal J}}

\def\cL{{\cal L}}
\def\cM{{\cal M}}

\def\cO{{\cal O}}

\def\cR{{\cal R}}

\def\cV{{\cal V}}
\def\cW{{\cal W}}

\def\ssA{{\scriptscriptstyle A}}
\def\ssB{{\scriptscriptstyle B}}

\def\ssH{{\scriptscriptstyle H}}

\def\ssL{{\scriptscriptstyle L}}
\def\ssM{{\scriptscriptstyle M}}
\def\ssN{{\scriptscriptstyle N}}
\def\ssP{{\scriptscriptstyle P}}
\def\ssQ{{\scriptscriptstyle Q}}
\def\ssR{{\scriptscriptstyle R}}

\def\KK{{\scriptscriptstyle KK}}

\def\BD{{\scriptscriptstyle BD}}

\newcommand{\bmat}{\left(\begin{array}}
\newcommand{\emat}{\end{array}\right)}

\def\-{\hphantom{-}}

\def\s2{\frac{1}{2}}

\def\IF{\relax{\rm I\kern-.18em F}}
\def\II{\relax{\rm I\kern-.18em I}}
\def\IP{\relax{\rm I\kern-.18em P}}
\def\IC{\relax{\rm I\kern-.48em C}}
\def\IR{\relax{\rm I\kern-.18em R}}
\def\IK{\relax{\rm I\kern-.20em K}}
\def\IM{\relax{\rm I\kern-.25em M}}

\def\y2{Y_{\ssM\ssN} Y^{\ssM\ssN}}
\def\Riem2{R_{\ssA\ssB\ssM\ssN} R^{\ssA\ssB\ssM\ssN}}
\def\Ricci2{R_{\ssM\ssN} R^{\ssM\ssN}}

\def\f2{F^{a}_{\ssM\ssN} F^{\ssM\ssN}_a}

\def\Asl{\hbox{/\kern-.7500em\it A}} % A slash
\def\dsl{\hbox{/\kern-.5500em$\partial$}}
\def\pxpsl{\hbox{/\kern-.5600em$p$}}
\def\Dsl{\,\raise.15ex\hbox{/}\mkern-13.5mu D}
\def \one{\relax{\rm 1\kern-.26em I}}

\def\exd{{\rm d}}

\def\nn{\nonumber}

\def\({\left(}
\def\){\right)}

\def\gR{g_\ssR}

\allowdisplaybreaks

\title{Gravitational Forces on a Codimension-2 Brane}

\author{C.P.~Burgess,${}^{1,2}$
L.~van Nierop${}^3$ and M.~Williams${}^{1,2}$\\

$^1$ Department of Physics \& Astronomy, McMaster University,
 Hamilton ON, Canada\\
$^2$   Perimeter Institute for Theoretical Physics,
 Waterloo ON, Canada\\
$^3$ Abdus Salam ICTP, Strada Costiera 11, Trieste 34014, Italy
}

\date{}
\abstract { We compute the gravitational response of six dimensional gauged, chiral supergravity to localized stress energy on one of two space-filling branes, including the effects of compactifying the extra dimensions and brane back-reaction. We find a broad class of exact solutions, including various black-brane solutions. Several approximate solutions are also described, such as the near-horizon geometry of a small black hole which is argued to be approximately described by a 6D Schwarzschild (or Kerr) black hole, with event horizon appropriately modified to encode the brane back-reaction. The general linearized far-field solutions are found in the 4D regime very far from the source, and all integration constants are related to physical quantities describing the branes and the localized energy source. The localized source determines two of these, corresponding to the source mass and the size of the strength of a coupling to a 4D scalar mode whose mass is parametrically smaller than the KK scale. At large distances the solutions agree with those of 4D general relativity, but for an intermediate range of distances (larger than the KK scale) the solutions better fit a Brans-Dicke theory. For a realistic choice of parameters the KK scale could lie at a micron, while the crossover to Brans-Dicke behaviour could occur at around 10 microns. While allowed by present data this points to potentially measurable changes to Newton's Law arising at distances larger than the KK scale.
}

\begin{document}

\section{Introduction}
\label{sec:Introduction}

If the vacuum has an energy density, it appears to gravitate much more weakly than do other energy sources \cite{Accelerating}. Attempts to understand how this might be possible have spawned many efforts towards modifying gravity \cite{ModGravRev, CCMod}, with (so far) disappointing results \cite{Wbgnogo, LesHouchesRev}. The broad features of what is required are clear: a Lorentz-invariant vacuum energy should largely not gravitate while ordinary mass distributions should do so unchanged from the predictions of general relativity (over solar-system distances). By modifying gravity one hopes to learn why these two types of energy sources should differ from one another in their gravitational response, with the difference possibly being traced to their energy distribution in space and/or time.

Higher-dimensional models provide a particularly concrete arena for exploring such modifications, for several reasons. First, although nonzero stress-energy must curve spacetime, when there are extra dimensions it is not necessarily true that the resulting curvature occurs within the four large dimensions visible to cosmologists. Second, if only one or two dimensions are as large as the micron-ish scales associated with the Dark Energy density --- which, remarkably, remains phenomenologically viable \cite{ADD, InvSqTests1, InvSqTests} --- then long-range forces within these dimensions do not fall off with distance. This makes it impossible to neglect the effects of back-reaction when computing the curvature produced by a given energy distribution, such as by the tension of various space-filling branes situated around the extra dimensions. Furthermore, the effects of back-reaction are comparatively poorly understood, and it is precisely in such poorly explored areas that one hopes something may have been missed by previous searchers.

These ideas have been explored in both 5D \cite{5Dmodels} and 6D \cite{6Dmodels} models, where it was shown that extra dimensions in themselves turn out not to be enough. Although it is true that back-reaction causes the low-energy 4D vacuum energy to differ significantly from the naive `probe' approximation, the difference doesn't appear to favour particularly small curvatures once all effects are taken into account. The situation seems more promising if there are two extra dimensions that are both large {\em and} supersymmetric \cite{Towards, SLEDrevs, TNCC}, not least because extra-dimensional supersymmetry can 
forbid an extra-dimensional cosmological constant. More generally, supersymmetry --- and the classical scale invariance that extra-dimensional supergravities generically enjoy --- can combine to help suppress both classical \cite{NoDilFlat, LargeDim} and quantum \cite{DistSUSY} contributions to the vacuum energy. In some cases the resulting observable 4D vacuum energy can be as small as the Kaluza-Klein (KK) scale, even when the scale of supersymmetry breaking within particle physics on a brane is much higher \cite{TNCC, DistSUSY}.

Back-reaction plays an important role in all of these results, with the tensions of the various branes turning out to be cancelled or partially cancelled by the curvatures the branes set up within the extra dimensions \cite{6Dmodels}. Among the issues that remain to be pinned down in these models is whether or not this same back-reaction also modifies how ordinary objects gravitate, in a way that is consistent with the abundance of tests of general relativity, and its many successful applications to cosmology.

In this paper we explicitly compute how point objects gravitate in supersymmetric large extra-dimensional (SLED) models, with the goal of filling in some of these missing steps. We do so using the specific example of 6D chiral, gauged supergravity \cite{NS}, though we expect many of our results to carry over to other 6D supergravities inasmuch as they largely rely on general symmetry properties. One hopes both to see whether the models are ruled out by existing measurements and, if not, to see if one can find robust observable differences from general relativity with which the proposals can be tested definitively. Our results extend earlier calculations that have been performed for non-supersymmetric 6D models \cite{PST}, as well as partial results that are known for supersymmetric 6D models \cite{Salvio}.

We make the following findings:
\begin{itemize}
\item We identify a broad class of exact classical solutions to the field equations that include explicitly both the compactification of the two extra dimensions and the back-reaction of up to two brane sources (whose properties need not be identical to one another). Besides the well-known maximally 4D-symmetric vacuum solutions, we also have a broad class of exact black-brane solutions \cite{blackbrane} (to which we argue the solution for generic point masses asymptotes at large distances).
\item For small enough black holes we identify how the supergravity solutions eventually approach the form of a black hole solution to pure Einstein gravity, provided they don't couple directly to the 6D dilaton of the 6D gravity supermultiplet. Brane back-reaction modifies these solutions \cite{Nemanja} much as does a black hole localized on a cosmic string in four dimensions \cite{AFV}.
\item We identify the linearized equations that govern the transition between the near-field 6D and far-field 4D solutions, and show that brane back-reaction does not alter the naive expressions for the conserved charges (like mass or angular momentum) that can be related in these solutions to the asymptotic fall-off towards the vacuum solution at large distances.
\item We obtain the general solution (for static point sources) to these linearized equations in the far-field regime for which the fields fall off as $1/r$, in the 4D manner. By imposing the near-brane boundary conditions we identify the physical interpretation of the integration constants in these solutions. We show that two constants survive undetermined by the near-brane boundary conditions, and must be fixed in terms of the properties of the point source.
\item We argue that these last two boundary conditions correspond to the two parameters that arise within the low-energy 4D theory that describes the low-energy 4D limit that is appropriate to the $1/r$ limit. Besides the massless 4D graviton, this 4D theory also includes a 4D scalar breathing mode that is not precisely massless, but its mass, $\mu$, is parametrically light compared with the KK scale: $\mu \simeq m_\KK \delta$, with $\delta \ll 1$ of order the defect angle of the back-reacting branes.  This scalar couples to brane matter like a Brans-Dicke scalar, and matching to the 6D linearized solution reveals the two integration constants to be the point-source mass and Brans-Dicke coupling (and so scalar charge).
\item Although the Brans-Dicke coupling found by dimensional reduction is large enough to have been ruled out, the Brans-Dicke solution only applies for $r \ll \mu^{-1}$. For distances larger than this the scalar field falls exponentially to zero and the far-field behaviour is precisely as expected for 4D general relativity, given the point source's mass. In the practical examples that are known to give acceptably large quantum-generated vacuum energies \cite{DistSUSY}, $m_\KK \simeq 0.1$ eV and $\delta \simeq 0.1$ and so $1/\mu \simeq 10$ $\mu$m. At this range the theory crosses over to a Brans-Dicke form, and so in particular would appear as a modification at these distances to the size of Newton's constant, rathern than a change to the inverse square law. (A crossover to a $1/r^4$ force law ultimately does occur, but only at the KK scale, $1/m_\KK \simeq 1$ $\mu$m.)
\end{itemize}

The remainder of this paper marshals our arguments as follows. First, \S\ref{sec:FEandsolns} sets up the field equations for the 6D supergravity of interest, and briefly describes the exact solutions that describe compactified vacuum and more general configurations, including brane back-reaction. This section describes, in particular, the 6D black holes that govern the near-field limit (for sufficiently small black holes) and the black branes that govern the far-field limit, as well as the scales where these approximate forms become valid. Next, \S\ref{sec:matching} provides the tools needed to relate the physical brane and point-source properties to the integration constants obtained from integrating the field equations. These come in two types: near-brane boundary conditions that capture the back-reaction of the branes; and expressions for conserved charges (like mass) that relate the asymptotic far-field solution to the source properties. \S\ref{sec:farfield} then linearizes the field equations and gives their general solutions in the far-field four-dimensional limit. The boundary conditions of \S\ref{sec:matching} are applied and used to eliminate all but two of the integration constants. The remaining two are fixed in terms of the source mass and `scalar charge' by comparing the far-field form of the 6D solutions with the far-field form of the corresponding solutions to the 4D effective field theory that should govern this limit. This is followed, in \S\ref{sec:gpheno}, by a discussion of the implications for short-distance tests of the Newtonian inverse-square law. Finally, \S\ref{sec:discussion} provides a preliminary discussion of some of the implications of these calculations.

\section{Field equations and some solutions}
\label{sec:FEandsolns}

In this section we set up and solve the 6D supergravity field equations in detail, including the effects caused by the back-reaction of the branes on their gravitational environments. After first briefly summarizing the back-reacted vacuum solutions, we then identify the higher-dimensional gravitational field (and other bulk fields) generated by a point mass localized on one of the branes.

\subsection{The field equations}

The 6D theory of interest is 6D chiral gauged supergravity \cite{NS}, whose action when truncated to just the graviton $g_{\ssM\ssN}$, an $R$-symmetry, $U(1)_\ssR$, gauge field, $F_{\ssM\ssN}$, and the dilaton $\phi$, is\footnote{We use a `mostly-plus' metric, and adopt Weinberg's curvature conventions \cite{GandC}.}
\be
 S = -\int \! \exd^6 x \sqrt{-g} \bigg[ \frac1{2\kappa^2} \Big( \cR + \pd_\ssM \phi \, \pd^\ssM\phi \Big) + \frac{1}{4 \gR^2} \, e^{-\phi} F_{\ssM\ssN} \, F^{\ssM\ssN} + \frac{2 \gR^2}{\kappa^4} \, e^\phi \bigg] \,,
\ee
where $\kappa^2 = M_g^{-4}$ is the 6D gravitational coupling and $\gR^2 \ll \kappa$ is the scale appearing in the 6D gauge coupling, $g^2(\phi) := \gR^2 \, e^{\phi}$, for the Maxwell field, $F_{\ssM\ssN}$.

The corresponding field equations are
\begin{gather} \label{NSFEs}
 \cR_{\ssM\ssN} + \pd_\ssM\phi\,\pd_\ssN\phi + \frac{\kappa^2}{\gR^2} \,  e^{-\phi} F_{\ssM\ssP} {F_\ssN}^\ssP - \frac12\left( \frac{\kappa^2}{4 \gR^2} \, e^{-\phi} F_{\ssP\ssQ} F^{\ssP\ssQ} - \frac{2\gR^2}{\kappa^2} \; e^\phi \right) g_{\ssM\ssN} = 0 \qquad\\
 \pd_\ssM \Big(\sqrt{-g} e^{-\phi} F^{\ssM\ssN} \Big) = 0 \\
 \square \phi + \left( \frac{\kappa^2}{4\gR^2} \, e^{-\phi} F_{\ssP\ssQ} F^{\ssP\ssQ} - \frac{2\gR^2}{\kappa^2} \; e^\phi \right) = 0 \,.
\end{gather}
Notice these equations are invariant under the classical rigid scaling symmetry under which $g_{\ssM\ssN} \to c \, g_{\ssM \ssN}$ and $e^{-\phi} \to c \, e^{-\phi}$, for arbitrary constant $c$.

\subsection{Two classes of exact compactified solutions}

In this section we describe two classes of exact solutions to the above field equations. Our main interest is in solutions to the above field equations arising from various matter sources. But before describing these it is worth briefly recapping the properties of vacuum solutions, for which the four on-brane directions are maximally symmetric.

\subsubsection*{Vacuum solutions}
\label{sec:vacsolutions}

We start here with the vacuum solutions, focussing on branes that do not couple to the 6D dilaton, $\phi$, since this is sufficient to ensure that maximally symmetric on-brane geometries are flat \cite{NoDilFlat, 6DdS} (making them the ones of most interest for applications to the cosmological constant problem \cite{TNCC}).

In this case the most general solutions involving two or fewer branes can be written in closed form \cite{GGP}, and are given by\footnote{We use here the coordinates of ref.~\cite{DistSUSY}, since these are more physically transparent for the present purposes than are those used in the earlier literature.}
\be \label{eq:GGPmetric}
 \exd s^2 = \cW^2(\theta) \, \exd \hat s^2_4 + a^2(\theta) \Big(\exd\theta^2 + \alpha^2(\theta) \sin^2\theta \,\exd\varphi^2 \Big) \,,
\ee
where the field equations imply $\exd \hat s^2_4 := \eta_{\mu\nu} \exd x^\mu \exd x^\nu$,
\be \label{eq:GGPa}
 a(\theta) =  a_0 \cW(\theta) \qquad \hbox{with} \qquad a_0 := \frac{\kappa\,e^{-\phi_0/2}}{2\gR} \,,
\ee
and
\be
 \alpha(\theta) = \frac\lambda{\cW^4(\theta)} \,,
\ee
with
\be
 \cW^4(\theta) = e^{\eta} \sin^2\frac{\theta}{2} + e^{-\eta} \cos^2 \frac{\theta}2
 = \cosh \eta - \sinh \eta \,\cos\theta \,.
\ee
The background gauge field (with flux quantum $n = \pm 1$) is given by
\be
 F_{\theta\varphi} = \pm \frac{\lambda \,\sin\theta}{2 \, \cW^8(\theta)}  \,,
\ee
and the dilaton profile is
\be
 e^{\phi(\theta)} = \frac{e^{\phi_0}}{\cW^2(\theta)} \,.
\ee

Here $\eta$, $\lambda$ and $\phi_0$ are three integration constants. Two of these can be related to brane properties by the near-brane boundary conditions \cite{HiCoDBCs}, as is seen most simply\footnote{Recall that the defect angle is related to the brane lagrangian density, $\cL_b = \sqrt{-\gamma} \; L_b$ by $\delta_b = \kappa^2 L_b/2\pi$ \cite{LargeDim}, which reduces to the standard expression \cite{Vilenkin} for a pure-tension brane (for which $L_b = T_b$).} if they are traded for the defect angles, $\delta_b = 2\pi(1 - \alpha_b)$, for the conical singularities due to the back-reaction of the branes located at the two poles (labeled by $b = \pm$). Defining $\alpha_+ := \alpha(\theta = 0)$, $\alpha_- := \alpha(\theta = \pi)$, we have
\be \label{eq:Wpmforms}
 \lambda = \sqrt{\alpha_+\alpha_-} \qquad \hbox{and} \qquad e^{\eta} = \sqrt{\frac{\alpha_+}{\alpha_-}}  \,.
\ee
The constant, $\phi_0$, labels the one-parameter class of solutions dictated by the classical scale invariance mentioned below eqs.~\pref{NSFEs}. For the particular special case $\alpha_+ = \alpha_-$ the function $\cW(\theta)$ (and so also $\phi(\theta)$, $a(\theta)$ and $\alpha(\theta)$) is constant and the geometry \pref{eq:GGPmetric} reduces to the simple rugby-ball solution \cite{SS, Towards}.

The extra-dimensional volume is
\be \label{V2result}
 \cV_2 = 2\pi \int_0^\pi \exd \theta \sqrt{\tilde g} = 8 \pi a_0^2 \left[ \frac{(\alpha_+ \, \alpha_-)^{3/4}}{\sqrt{\alpha_+} + \sqrt{\alpha_-}} \right] \,,
\ee
and the 6D and 4D Planck scales, $M_g$ and $M_p$, are related by
\be
 M_p^2 = 2\pi M_g^4 \int_0^\pi \exd \theta \sqrt{\tilde g} \; \cW^4 = 4 \pi a_0^2 M_g^4 \, \sqrt{\alpha_+ \, \alpha_-} \,,
\ee
where the (rationalized) 4D Planck mass, $M_p$, is defined relative to the 4D Newton constant, $G_4$, by $8 \pi G_4 = M_p^{-2}$. Notice that these reduce to the spherical case, $\cV_2 = 4\pi a_0^2$ and $M_p^2 = \cV_2 M_g^4$, when $|\alpha_\pm - 1| \simeq \cO(\kappa^2 T_\pm/2\pi) \ll 1$, which is the regime of most practical interest.

\subsubsection*{More general solutions and black branes}
\label{sec:Action}

The above vacuum solutions immediately generalize to a broader class of exact compactified solutions. The idea behind this generalization is the observation that new exact solutions to eqs.~\pref{NSFEs} can be extracted given any known solution for which the 4 large dimensions are maximally symmetric.

To this end consider a metric of the simple warped-product form
\begin{equation} \label{eq:warpedproduct}
 \exd s^2 = \cW^2(y) \, \hat g_{\mu\nu}(x) \, \exd x^\mu \exd x^\nu + \tilde g_{mn}(y) \, \exd y^m \exd y^n \,,
\end{equation}
where $x^\mu$ (with $\mu = 0,1,2,3$) and $y^m$ (with $m = 4,5$) are respectively 4D and 2D coordinates. When the 4D metric, $\hat g_{\mu\nu}$, is maximally symmetric this form includes a great many known solutions, including the original spherical Salam-Sezgin solution \cite{SS} and its generalizations to include branes. These include the solutions with two (or fewer \cite{TearDrop}) branes described above \cite{GGP, NoDilFlat}, as well as multiple-brane configurations \cite{MultiBranes} --- for all of which $\hat g_{\mu\nu}$ is flat. It also includes solutions where $\hat g_{\mu\nu}$ is curved ({\em e.g.} 4D de Sitter) \cite{6DdS}.

Our present interest is in solutions that are {\em not} maximally symmetric in the 4D directions. For these purposes we now prove the following proposition: given any solution of the form of eq.~\pref{eq:warpedproduct} for which the 4D geometry is maximally symmetric, there is a broad family of other exact solutions for which the 4D metric is not maximally symmetric. In particular, the new solutions are also described by eq.~\pref{eq:warpedproduct}, with precisely the same functions $\cW(y)$ and $g_{mn}(y)$, but with the maximally symmetric geometry, $\hat g_{\mu\nu}$ replaced by any $y^m$-independent Einstein metric,\footnote{An Einstein metric satisfies $R_{\mu\nu} = k \, g_{\mu\nu}$, for constant $k$.} $g_{\mu\nu}$, whose curvature scalar agrees with the original maximally symmetric solution: $R = \hat R$.

This type of new solution exists because the 6D Ricci curvature, $\cR_{\ssM\ssN}$, associated with the above metric ansatz can always be expressed as
\ba
 \cR_{mn} &=& \tilde R_{mn} + 4\left(\pd_m \pd_n W + \pd_m W \pd_n W - \tilde\Gamma^p_{mn}\pd_p W\right)\nn\\
 \cR_{\mu\nu} &=& \hat R_{\mu\nu} + \left[\pd_m( \tilde g^{mn}\pd_n W) + 4 \tilde g^{mn}\pd_m W \pd_n W + \tilde g^{mn}\tilde\Gamma^p_{mp}\pd_n W \right] \hat g_{\mu\nu}\nn\\
 \cR_{m\mu}&=&0
\ea
where $W := \ln \cW$ and quantities with a tilde are constructed just from the metric $\tilde g_{mn}$ while hatted quantities are constructed with $\hat g_{\mu\nu}$. The basic observation is that the field equations depend on $\hat g_{\mu\nu}$ only through the combination $\hat R_{\mu\nu} = \frac14 \, \hat R \, \hat g_{\mu\nu}$, and so are exactly the same as they would be for a maximally symmetric space provided $\hat R$ is constant.

In particular, for each 4D flat solution of the form \pref{eq:warpedproduct} (such as the ones described in \S\ref{sec:vacsolutions}\ above), another equally good solution is obtained if $\hat g_{\mu\nu}(x)$ is instead an arbitrary 4D Ricci-flat metric, with all other functions (like $\cW$, $\phi$, {\em etc}.) unchanged. Similarly, for each 6D solutions with a 4D de Sitter geometry there is a new class of exact solutions for which $\hat R_{\mu\nu} = \frac14 \hat R \hat g_{\mu\nu} \ne 0$, with all other functions unchanged. This result allows the immediate identification of asymptotically flat (or de Sitter) black branes: first solve the vacuum field equations for a flat (or de Sitter) brane, and then replace the 4D metric by an asymptotically flat (or de Sitter) black-hole spacetime.

As stated earlier, our present interest is in solutions for which any brane sources do not couple to the 6D dilaton, $\phi$, since for 4D maximally symmetric geometries this suffices to ensure $\hat g_{\mu\nu}$ is flat \cite{NoDilFlat, 6DdS}. For any such solution we have just seen that another exists where $\hat g_{\mu\nu}$ is instead an arbitrary 4D Ricci-flat metric, and so in particular $\hat g_{\mu\nu}$ could be given by the Schwarzschild or Kerr metric. For instance, the Schwarzschild black brane is given explicitly by
\be \label{eq:GGPmetric}
 \exd s^2 = \cW^2(\theta) \, \exd \hat s^2_4 + a^2(\theta) \Big(\exd\theta^2 + \alpha^2(\theta) \sin^2\theta \,\exd\varphi^2 \Big) \,,
\ee
where the functions $\cW$, $a$, $\alpha$, $\phi$ and $F_{mn}$ are as given in \S\ref{sec:vacsolutions}, but with the 4D geometry given by
\be
 \exd \hat s^2_4 := \hat g_{\mu\nu} \exd x^\mu \exd x^\nu = - f(r) \exd t^2 + \frac{\exd r^2}{f(r)} + r^2 \Bigl( \exd \xi^2 + \sin^2 \xi \, \exd \zeta^2 \Bigr) \,,
\ee
with $f(r) = 1 - r_s/r$ and $r_s = 2 \, G_4 M$ denoting the 4D Schwarzschild radius.

As is verified explicitly below, this black brane solution should agree asymptotically with the gravitational field outside of a black hole located on one of the branes, at distances much further than both the Schwarzschild and the KK scales; $r \gg r_s$ and
\be
 r \gg a_0 = \frac{\kappa\,e^{-\phi_0/2}}{2\gR} \,.
\ee
This is true in particular for the distances relevant to astrophysical black holes. Physically, the far-field limit of a point source on a brane agrees with the far-field black brane solution because at large distances only the lowest multipole is relevant, and this cannot resolve where the mass is situated within the extra dimensions.

\subsection{Some relevant scales}

When discussing the gravitational field of a source of mass $M$ localized on a brane, there are three important scales that govern the limits of most physical interest. These are:
\begin{itemize}
\item The KK scale, $a_0$, defined in eq.~\pref{eq:GGPa}, that sets the size of the extra dimensions;
\item The 4D Schwarzschild radius, $r_s = 2 G_4 M = M/(4\pi M_p^2)$, that sets the size of the event horizon of a 4D Schwarzschild black hole of mass $M$;
\item The 6D Schwarzschild radius,\footnote{We use $\rho$ to denote the 6D radial coordinate, keeping $r$ as the radial coordinate in the 4 dimensions parallel to the branes.} $\rho_s^3 = 3 \kappa^2 M/(4\pi)^2 = 3 M/(4\pi M_g^2)^2$, that sets the size\footnote{See Appendix \ref{app:newtonian} for a justification of the numerical factors in $\rho_s$.} of the event horizon of a 6D Schwarzschild black hole of mass $M$.
\end{itemize}
Notice that these scales are related to one another,
\be
 r_s = \frac{M}{4\pi M_p^2} \simeq \frac{M}{(4\pi a_0)^2 M_g^4}
 = \frac{\rho_s^3}{3 a_0^2} \,,
\ee
and so $\rho_s = r_s$ when $\rho_s^2 \simeq 3 a_0^2$. Writing
\be
 r_s = \left( \frac{M}{M_0} \right) \, \frac{a_0}3 \qquad \hbox{and} \qquad
 \rho_s = \left( \frac{M}{M_0} \right)^{1/3} a_0 \,,
\ee
we see $r_s \gsim \rho_s \gsim a_0$ for $M \gsim M_0$, and for $M \lsim M_0$ we have $r_s \lsim \rho_s \lsim a_0$. For $a_0$ of order a micron the transitional mass corresponds to $M_0 := a_0/G_4 \simeq M_\odot/(3 \times 10^9) \simeq 7 \times 10^{20}$ kg (about 1\% of the mass of the Moon).

We next consider several examples of how these scales control the relative size of physical effects for gravitational geometries.

\subsubsection*{Black brane instability}

For 6D Einstein gravity, for which $\cR_{\ssM\ssN} = 0$, relatively general considerations \cite{GregoryLaflamme} argue that the black-brane solution becomes unstable when $M \ll M_0$, and so for which $r_s \ll \rho_s \ll a_0$. In this limit an instability arises for extra-dimensional modes of wavelength $\lambda \simeq \rho_s$, which like to fragment the black brane into 6D black holes.

We expect a similar phenomenon also to occur for the solutions of eqs.~\pref{NSFEs}, in which case the black-brane solution discussed above should only be appropriate in the $M \gg M_0$ limit for which $r_s \gg \rho_s \gg a_0$. Black branes this large escape the instability because the unstable modes are too large to fit into the extra dimensions, since $\rho_s \gg a_0$.

In the other limit, when $M \ll M_0$, a more 6D black-hole solution should apply. Although the precise solutions appropriate for the 6D black hole in this limit are not known for eqs.~\pref{NSFEs}, including both brane back-reaction and extra-dimensional compactification, we extract below many of its features in physically interesting limits. In particular, we argue that close enough to the event horizon it is the response to the presence of the source mass $M$ that dominates the geometry, and this is approximately described (assuming the black hole is not itself a source for the bulk dilaton and Maxwell fields\footnote{When the branes {\em do} couple to $\phi$ and $A_\ssM$ the near-brane asymptotics can instead be captured by using the approximate Kasner-type near-brane solutions identified in \cite{6DdS}.}) by a standard 6D Schwarzschild black hole, with slightly modified event horizon where this is intersected by the brane. Further from the black hole (but still less than the KK scale) the energy density of the other bulk supergravity fields becomes important and so governs the transition from 6D to 4D behaviour.

\subsubsection*{6D black hole vs 6D KK black hole}

The region wherein the 6D black hole solution to eqs.~\pref{NSFEs} is well approximated by a standard 6D black hole is $\rho \lsim \rho_\star$, with $\rho$ denoting 6D radius (defined more precisely below) and $\rho_\star$ defined by the distance for which the energy density of the bulk fields $\phi$ and $F_{\ssM\ssN}$ become important. It is only once $\rho \gsim \rho_\star$ that the solution `learns' that it sits within compact higher dimensions and so modifies its long-distance behaviour. The same statement also applies, of course, to the exterior gravitational field of a source that is much smaller than the extra dimensions but is not a black hole because its physical size is larger than $\rho_s$.

To estimate $\rho_\star$ we assume the energy density of $\phi$ and $F_{\ssM\ssN}$ to be the same order as for the vacuum spacetime of \S\ref{sec:vacsolutions}, and so the $F^2$ and $e^\phi$ terms in the 6D Einstein equation are of order $1/a_0^2$ in size. But in 6D the components of the Riemann tensor at distance $\rho$ from a point mass are of order $\rho_s^3/\rho^5$, and so the generic curvature components are much larger than the components of the bulk stress energy for any
\be
 \rho^5 \lsim \rho_\star^5 \simeq \rho_s^3 a_0^2 \,, \qquad \hbox{and so} \qquad
 \rho_s \lsim \rho_\star \lsim a_0 \,,
\ee
where the last inequalities use $M \lsim M_0$ (and so $\rho_s \lsim a_0$), as is required for the 6D black-hole solution to be relevant. When $\rho \ll \rho_\star$ the contributions of the bulk stress energy should be negligible and the near-horizon components of the metric should dominantly be controlled by the vacuum field equations, $\cR_{\ssM \ssN} \simeq 0$.

\subsection{Approximate 6D black holes}

For black holes much smaller than the KK size, the above arguments suggest the black-hole solution to eqs.~\pref{NSFEs} is well approximated by a standard 6D Schwarzschild black hole \cite{HiDBH, ADDBHrev},
\be \label{6DBHsph}
 \exd s^2 = - h(\rho) \exd t^2 + \frac{\exd \rho^2}{h(\rho)} + \rho^2 \exd \Omega_4 \,,
\ee
where $h(\rho) = 1 - (\rho_s/\rho)^3$ and $\exd \Omega_4$ is the line-element on the unit 4-sphere. Here $\rho_s$ is the 6D event horizon, related to the black-hole mass as stated earlier: $\rho_s^3 \propto G_6 M$.

Although this solution should be adequate at most places near the horizon, it must break down where the horizon intersects the brane world sheet, which we can take to be at the poles of the 4-sphere. The boundary conditions at these points can be implemented following the codimension-two brane surgery described in \cite{Nemanja}, adapted from the string/black-hole solution of \cite{AFV}, in which a wedge having the appropriate defect angle is removed from the 4-sphere geometry. A set of coordinates for the resulting metric that is convenient for our later purposes is given by \cite{Nemanja}
\be \label{6DBHcyl1}
 \exd s^2 = - \left( \frac{4 \varrho^3 - \rho_s^3}{4 \varrho^3 + \rho_s^3} \right)^2
 \exd t^2 + \left( 1 +  \frac{\rho_s^3}{4\varrho^3} \right)^{4/3} \exd s_5^2 \,,
\ee
where
\be \label{6DBHcyl2}
 \exd s_5^2 = \exd r^2 + r^2 \Bigl( \exd \xi^2 + \sin^2 \xi \, \exd \zeta^2 \Bigr) + a_0^2 \Bigl( \exd \theta^2 + \alpha^2 \theta^2 \, \exd \varphi^2 \Bigr) \,,
\ee
is the flat metric on a 5D cone with $\alpha = 1 - \delta/2\pi$ measuring the defect angle. Here $\varrho$ is related to the other coordinates by $\varrho^2(r, \theta) = r^2 + a_0^2 \theta^2$, and to the 6D Schwarzschild radial coordinate by
\be
 \rho = \varrho \left( 1 + \frac{\rho_s^3}{4 \varrho^3} \right)^{2/3}
 \simeq \varrho + \frac{\rho_s^3}{6 \varrho^2} + \cdots \,,
\ee
where the approximate equality assumes both $\rho$ and $\varrho$ to be much larger than $\rho_s$.

This solution should be a good approximation for $\rho_s < \rho \ll \rho_\star < a_0$, while for distances $\rho \gsim \rho_\star$ the stress energy of the bulk fields becomes important, allowing the simple Schwarzschild solution of this section to cross over at the KK scale to a solution with 4D $1/r$ asymptotics. We return to an explicit construction of some of the features of this solution in \S\ref{sec:farfield} below.

\section{Matching conditions}
\label{sec:matching}

Before turning to a more general solution to the linearized field equations, we first digress to describe the boundary conditions that must be applied to relate integration constants to the physical choices made on the branes. These boundary conditions come in two types: those applied near the branes but far from the position of the point source, and those that capture the flux of energy and angular momentum of the point source.

\subsection{Near-brane limits}

We start with the near-brane boundary conditions that apply far from the position of the point source. These have the same form as they would have in the absence of the point source, and so also apply to the vacuum solutions described earlier.

How far from the point mass must we be to neglect its effects on the near-brane boundary conditions? To estimate this we ask how far on the brane we must be from the source to have its energy density fall below the energy density set by the tension, $T$, of the brane. This occurs when $M \simeq 4 \pi r_\star^3 T$, and so
\be
 r_\star^3 \simeq \frac{M}{4 \pi T} \gsim \frac{M}{2 M_g^4} = 2\pi \rho_s^3 \,,
\ee
where the inequality uses $\kappa^2 T/2\pi \lsim 1$, which is the domain for which our semi-classical methods work best. For the case of most interest numerically $T \simeq (5 \; \hbox{TeV})^4$ and $M_g \simeq 10$ TeV, so we see $\kappa^2 T/2\pi \simeq 1/32\pi \sim 10^{-2}$ and $r_\star \simeq 2^{4/3} \rho_s \simeq 2.5 \rho_s$ is not much larger than $\rho_s$.

For $r \gg r_\star$ the near-brane limit has the form found in refs.~\cite{Cod2Matching}. Assuming the brane action of the form
\be
 S_b = - \int \exd^4 x \sqrt{- \gamma} \; L_b  \,,
\ee
with $\gamma_{\mu\nu}$ the induced metric as before, and where $L_b$ is independent of $\phi$, the near-brane dilaton boundary condition becomes
\be
 \lim_{\theta \to \theta_b} \sqrt{-h} \; g^{\ssM\ssN} \, n_\ssM \partial_\ssN \phi = 0 \,,
\ee
where $h_{ab}$ is the induced metric on surfaces of constant $\theta$ and $n_\ssM$ is the corresponding normal vector; the brane is assumed to be situated at $\theta = \theta_b$, with $b = \pm$. (For the vacuum solutions described earlier there are two branes with $\theta_+ = 0$ and $\theta_- = \pi$.)

Similarly, the near-brane limit of the metric satisfies
\be \label{eq:metmatch}
 \lim_{\theta\to\theta_b} \sqrt{-h} \Bigl[ \tfrac12 \big(K^{ab} - K \, h^{ab}\big) - (\hbox{flat}) \Bigr] = -\frac{\kappa^2}{2\pi} \left( \frac{\delta S_b}{\delta h_{ab}} \right) \,,
\ee
where $K_{ab}$ is the extrinsic curvature for the surfaces of fixed $\theta$, $K = g^{ab} K_{ab}$ and `flat' denotes the same expression evaluated in the absence of a brane. (In practice eq.~\pref{eq:metmatch} is most useful for the on-brane directions, for which $h_{\mu\nu} = \gamma_{\mu\nu}$, since it is only the dependence on the 4D metric, $\gamma_{\mu\nu}$, that is easily specified for a codimension-2 brane. The remaining components of $K^{ab}$ are instead determined using the `Hamiltonian constraint' for evolution in the $\theta$ direction \cite{HiCoDBCs}.) 

The gauge field satisfies a related constraint relating $n_\ssM F^{\ssM a}$ to $\delta S_b/\delta A_a$. Physically, this relates the magnetic flux threading a curve, $\Gamma$, encircling the brane,
\be \label{eq:Phibdef}
 \Phi_b = \lim_{\theta \to \theta_b} \, \oint_\Gamma A_\ssM \exd x^\ssM  \,,
\ee
to parameters appearing within the brane action. (For instance, in terms of the constant, $\cC_b$, appearing in the action, \pref{eq:fluxq}, matching of the gauge-field at the brane implies the brane-localized flux is given by $\Phi_b = g_\ssR^2 \cC_b e^{\phi_b} $, where $\phi_b = \lim_{\theta \to \theta_b} \phi$. In practice, for branes localized at the two poles, we regard $\Phi_b$ as being given and use eq.~\pref{eq:Phibdef} to fix the near-brane limit of the gauge potential:
\be
 \lim_{\theta \to \theta_b} A_\varphi = b\,\frac{\Phi_b}{2\pi} = b\,\frac{\gR^2 \cC_b e^{\phi_b} }{2\pi} 
\ee
where, as before, $b = \pm 1$ labels the branes (and accounts for the difference in handedness as $\Gamma$ encircles each brane).

Specialized to the following diagonal metric
\be \label{eq:diagmetric}
 \exd s^2 = - e^{2A} \exd t^2 + e^{-2 C} \exd r^2 + r^2 e^{2W} \Bigl( \exd \xi^2 + \sin^2 \xi \, \exd \zeta^2 \Bigr) + a_0^2 \Bigl(e^{2E} \exd \theta^2 + e^{2B} \exd \varphi^2 \Bigr) \,,
\ee
and to a brane stress energy
\be
 T^{\mu\nu}_b = \frac{2}{\sqrt{-g}} \, \frac{\delta S_b}{\delta \gamma_{\mu\nu}} = - L_b \, \gamma^{\mu\nu} \,,
\ee
the above matching conditions imply the following near-brane limits
\be \label{branematch1}
 \lim_{\theta \to \theta_b} e^{B-E} \partial_\theta \phi =  \lim_{\theta \to \theta_b} e^{B-E}\partial_\theta A = \lim_{\theta \to \theta_b} e^{B-E}\partial_\theta C  = \lim_{\theta \to \theta_b} e^{B-E}\partial_\theta W = 0
\ee
and 
\be \label{branematch2}
 \lim_{\theta \to \theta_b} \Bigl( b\,e^{-E} \partial_\theta e^B - 1 \Bigr) = -\frac{\kappa^2 L_b}{2\pi} \,.
\ee
(The $b$ in front of $\partial_\theta e^B$ keeps track of whether or not $\exd \theta$ is an outward- or inward-pointing direction at the brane in question.) These are satisfied by the vacuum solutions of \S\ref{sec:vacsolutions}, and in particular the last of them relates the defect angle, $\delta_\pm$, at $\theta_+ = 0$ and $\theta_- = \pi$ to the brane action:
\be
\delta_b:=2\pi(1-\alpha_b) = \kappa^2 L_b \,.
\ee

\subsection{Conserved charges (like mass)}

We must also impose boundary conditions that relate the metric integration constants to the properties of the point mass. To do this we use a general expression that relates the mass of the source to the far-field asymptotic form of the its gravitational field \cite{ADM}. To this end we seek an expression for the mass (and other conserved quantities) that is similar to the Gauss' Law expression relating electric charge to asymptotic electric fields:\footnote{This is just the covariant version of the familiar integration: $Q = \int \sigma\, \exd V  = \int  \nabla \cdot E \, \exd V = \oint n \cdot E \, \exd S$.}
\be
 Q = \int_\Sigma \exd \Sigma_\ssM J^\ssM = \int_\Sigma \exd \Sigma_\ssM \, \nabla_\ssN F^{\ssM \ssN} = \frac12\oint_{\partial \Sigma} \exd \Sigma_{\ssM \ssN} F^{\ssM\ssN} \,,
\ee
where $\Sigma$ is a time-like hypersurface with asymptotic boundary $\partial \Sigma$, with respective measures $\exd \Sigma_\ssM$ and $\exd \Sigma_{\ssM \ssN}$. The last equality uses Stokes' Theorem to write the integral over the total derivative as a surface term.

%\ignore
\FIGURE[h!]{
  \centering
	\includegraphics[width=0.56\textwidth]{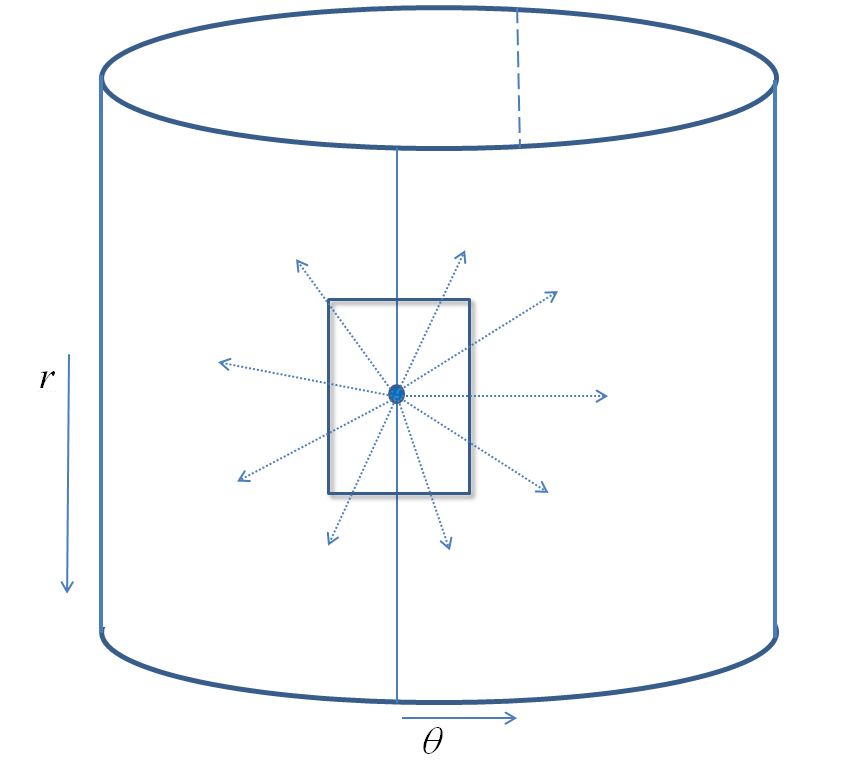}
    \caption{A sketch of the Gaussian pillbox enclosing the point mass localized on a brane, including outgoing gravitational flux lines. Two coordinates are drawn; the radial direction, $r$, along the brane, and the coordinate, $\theta$, along the extra dimensions. The antipodal brane is also shown as a dashed line.}
\label{fig:cylinder}
}
%\endignore

There is a general formalism for doing this within general relativity, whose features are briefly summarized in Appendix \ref{app:conservedcurrents}. The construction relies on the fields of interest being asymptotic to other fields that enjoy a symmetry. In the present instance we imagine the field far from a point source to asymptote at large distances to one of the vacuum solutions discussed in \S\ref{sec:vacsolutions}: $\phi \to \bar \phi$, $g_{\ssM \ssN} \to \bar g_{\ssM \ssN}$ and so on. Furthermore, this asymptotic vacuum solution has symmetries ({\em e.g.} time-translation and rotational invariance) and so admits Killing vector fields, $V^\ssM$, that satisfy
\be
 \ol \nabla_\ssM \ol V_\ssN + \ol\nabla_\ssN \ol V_\ssM = 0 \,,
\ee
where $\ol V_\ssM := \bar g_{\ssM \ssN} V^\ssN$.

The main result \cite{ADM, AbbottDeser} states that for each such a Killing vector there is a corresponding conserved charge, $Q[V]$, that can be written totally in terms of the asymptotic forms of the fields. As applied to time-translation invariance, $\ol V^\ssM \partial_\ssM = \partial_t$, this result allows the conserved mass to be written as an integral over a constant-$t$ slice,
\be \label{eq:microM}
 M  = \frac{1}{\kappa^2} \int \exd^5x \sqrt{\bar g} \; {\delta \cE^t}_t \,,
\ee
where ${\cE^\ssM}_\ssN = {\cG^\ssM}_\ssN + \kappa^2 {T^\ssM}_\ssN = 0$ is the 6D Einstein equation, and ${\delta \cE^\ssM}_\ssN$ is its linearization about the asymptotic background metric. $M$ defined this way is conserved, in the sense that it doesn't matter which $t$-slice is used in its evaluation. Furthermore, general arguments ensure the integrand on the right-hand side is a total derivative, and so can be expressed in terms of the asymptotic values of the fields.

To make this concrete let us specialize to the diagonal metric of eq.~\pref{eq:diagmetric}, and take the far-field solution to be a rugby ball solution of \S\ref{sec:vacsolutions} (for which $\alpha_+ = \alpha_- := \alpha$). In this case $A = \ol A + \delta A(r, \theta)$, $C = \ol C + \delta C(r, \theta)$, $W = \ol W + \delta W(r, \theta)$, $E = \ol E + \delta E(r, \theta)$, $B = \ol B + \delta B(r, \theta)$, $\phi = \ol \phi + \delta \phi(r, \theta)$ and $A_\ssM = \bar A_\ssM + \delta A_\ssM(r, \theta)$, with $\ol A = \ol C = \ol W = \ol E = 0$, $e^{\ol B} = \alpha \sin \theta$, $\ol\phi = \phi_0$ and $\ol F_{\theta\varphi} = \pm \frac12 \, \alpha \sin \theta$. A straightforward calculation (see Appendix \ref{app:conservedcurrents} for details) then reveals
\bea
 \sqrt{- \bar g} \; {\delta \cE^t}_t &=& \sqrt{- \bar g} \, \big({\delta \cG^t}_t + \kappa^2 {\delta T^t}_t\big) \nn\\
 &=& \frac{1}{a_0^2} \, \pd_\theta \left\{ \sqrt{-\bar g} \left[ \pd_\theta \Bigl( \delta C - \delta B - 2 \delta W \Bigr) + \Bigl(\delta E - \delta B \Bigr) \cot\theta  - \frac{\delta A_\varphi}{\ol F_{\theta\varphi}}\right] \right\} \nn\\
&&\quad -\pd_r \left\{ \sqrt{-\bar g} \left[ \pd_r \Bigl( \delta E + \delta B + 2 \delta W \Bigr) + \frac{2}r \, \Bigl(\delta C + \delta W \Bigr) \right] \right\} \,,
\eea
which is a total derivative, as promised.

Integrating over a Gaussian pillbox, as illustrated in figure \ref{fig:cylinder}, gives the following relation between the gravitational field threading the pillbox as a function of the mass enclosed:
\be \label{MisFrFth}
 M = \frac{1}{\kappa^2} \int_\Sigma \sqrt{\bar g} \; {\delta \cE^t}_t \, \exd^5 x = \cF_r + \cF_\theta \,,
\ee
where $\cF_r$ is the flux through the surface at fixed $r$ while $\cF_\theta$ is the flux through the surface at fixed $\theta$. Explicitly, using the above expressions, these fluxes become
\be
 \cF_r =  - \frac{8\pi^2 \alpha \, a_0^2}{\kappa^2} \int_{\theta_1}^{\theta_2} \exd \theta \left\{ r^2 \sin \theta \left[ \pd_r \Bigl( \delta E + \delta B + 2 \delta W \Bigr) + \frac{2}{r} \, \Bigl(\delta C + \delta W \Bigr) \right] \right\}_{r=0}^{r=r_\star} \,,
\ee
while the flux through the surfaces at $\theta = \theta_1$ and $\theta = \theta_2$ are
\be
 \cF_\theta = \frac{8\pi^2 \alpha}{\kappa^2} \int_0^{r_\star} \exd r \left\{ r^2 \sin \theta \left[ \pd_\theta \Bigl( \delta C - \delta B - 2 \delta W \Bigr) + \Bigl(\delta E - \delta B \Bigr) \cot\theta  - \frac{\delta A_\varphi}{\ol F_{\theta\varphi}}\right] \right\}_{\theta=\theta_1}^{\theta=\theta_2}\,.
\ee
Appendix \ref{app:conservedcurrents} shows how these formulae, evaluated using eqs.~\pref{6DBHcyl1} and \pref{6DBHcyl2} agree with the usual result for a black-hole mass evaluated using eq.~\pref{6DBHsph} evaluated on a 6D spherical surface, provided that the surface is chosen to lie close enough to the source (as is required for the 6D black hole to approximate a solution to the 6D supergravity equations).

For applications to fixing integration constants we wish to use eq.~\pref{MisFrFth} in the far-field regime, much further from the source than the size of the extra dimensions. In this case we can choose the surfaces of fixed $\theta$ to lie at the positions of the two branes, and use there the near-brane matching conditions described above. Since the energy density of the source does not compete with the much larger tension of the brane (with the possible exception of very relativistic sources that are smaller than $r_\star$ in size), both the asymptotic background and the solution in the presence of the source satisfy the same near-brane matching conditions. Consequently the linearized deviations must satisfy
\be \label{dbranematch}
 \lim_{\theta \to \theta_b} \sin \theta \, \partial_\theta \phi =  \lim_{\theta \to \theta_b} \sin \theta \, \partial_\theta A = \lim_{\theta \to \theta_b} \sin \theta \, \partial_\theta C  = \lim_{\theta \to \theta_b} \sin \theta \, \partial_\theta W = \lim_{\theta \to \theta_b} \sin \theta \, \delta A_{\varphi} = 0 \,.
\ee
The $\delta B$ condition is a bit more subtle given that $\partial_\theta \ol B \ne 0$, leading to
\be \label{dbranematch2}
 \lim_{\theta \to \theta_b} \sin \theta \Bigl[ (-\delta E + \delta B) \cot \theta + \partial_\theta \delta B \Bigr] = 0 \,.
\ee
Eqs.~\pref{dbranematch} and \pref{dbranematch2} imply $\cF_\theta = 0$, leaving only the flux at large $r$:
\be \label{eq:Mvslarger}
 M = \cF_r =  - \frac{8\pi^2 \alpha \, a_0^2}{\kappa^2} \int_{0}^{\pi} \exd \theta \left\{ r^2 \sin \theta \left[ \pd_r \Bigl( \delta E + \delta B + 2 \delta W \Bigr) + \frac{2}{r} \, \Bigl(\delta C + \delta W \Bigr) \right] \right\} \,.
\ee

Specializing this to a far-field regime where all perturbations are proportional to $1/r$ (more about when this is possible below), $\delta B = B_1/r$, $\delta C = C_1/r$, $\delta E = E_1/r$ and $\delta W = W_1/r$, this becomes
\be \label{eq:Mvslarger1}
 M = - \frac{8\pi^2 \alpha \, a_0^2}{\kappa^2} \int_{0}^{\pi} \exd \theta \sin \theta \Bigl(  2\, C_1 -  E_1 - B_1  \Bigr) \,.
\ee
This can be used to relate one combination of integration constants to the mass of the source. For example, for the black brane solution described earlier $E_1 = B_1 = 0$ while $2\,C_1 = 2\, A_1 = - r_s$ is $\theta$-independent, so
\be
 M = \frac{16\pi^2 \alpha \, a_0^2 \, r_s}{\kappa^2} = \frac{4 \pi \, r_s}{\kappa_4^2} = \frac{r_s}{2\,G_4} \,,
\ee
as expected.

\section{General linearized far-field solutions}
\label{sec:farfield}

In order to identify the possible integration constants to be determined in this way, we next obtain a more systematic calculation of the asymptotic behaviour of solutions in the far-field zone, much further from the source than the KK scale.

\subsection{Linearized solutions}

We now turn to the form of the long-range component of the fields in the interval $r \gg a_0$. Although this will include as a special case the exact $M > M_0$ black brane solutions of the previous section, it should also include the far-field limit of the more general (but more poorly understood) case of a localized point mass. This section summarizes the results, while Appendix \ref{Setup} provides more explicit details.

As above (though with $E=0$, a choice that can be made without loss of generality to leading order in $1/r$) we start with the metric {\it ansatz}
\bea
 \exd s^2 &=& a_0^2 \exd \theta^2 + e^{2B(\theta,r)} \,\exd\varphi^2 + \exd s_4^2 \\
 \exd s_4^2 &=& - e^{2A(\theta,r)} \, \exd t^2 + e^{-2C(\theta,r)} \, \exd r^2+r^2 e^{2W(\theta,r)} (\exd \xi^2 + \sin^2\xi \, \exd \zeta^2 ) \,,
\eea
and for simplicity we linearize the 6D field equations about the rugby-ball solution, which corresponds to
\bea
 &&\ol A = \ol C = \ol W = 0 \,, \quad
 e^{\ol B} = \alpha \, a_0 \sin \theta
 \,, \nn\\
  &&\ol\phi = \phi_0 = \ln\left(\frac{\kappa^2}{4g^2a_0^2}\right)
 \quad \hbox{and} \quad \ol F_{\theta \varphi} =  \frac{\alpha n}{2} \, \sin \theta \,,
\eea
with $n = \pm 1$.

Flux quantization also fixes the value of $\phi_0$ to
\be \label{fqrb}
 e^{\phi_0} = \frac{\delta}{2\gR^2 |\cC|}  = \frac{\kappa^2 T}{2 \gR^2 |\cC|} \,,
\ee
where the two (identical) brane actions are given by\footnote{The `brane-localized flux' term, proportional to $\cC_b$, may be written more covariantly as the integral of the 6D Hodge dual, ${}^\star F$, over the 4D brane world volume.}
\be \label{eq:fluxq}
 \cL_b = - \sqrt{- \gamma} \; \left[ T_b - \frac{\cC_b}{2} \, \epsilon^{mn} F_{mn} \right] + \cdots \,,
\ee
where the ellipses denote terms involving two or more derivatives. The constants $T_b$ (brane tension) and $\cC_b$ (brane-localized flux) are the quantities that specify the brane properties most relevant at low energies. Notice that fixing $\phi_0$ also fixes the size of the extra-dimensional geometry, through eq.~\pref{eq:GGPa}.

As shown in the appendix, linearization of eqs.~\pref{NSFEs} leads to eight independent linear field equations, given explicitly there as equations \pref{app:eq1} through \pref{app:eq8}. We seek solutions to these equations in the far-field regime corresponding to distances much further than the KK scale: $r \gg a_0$. Since this is well within the 4D regime, we seek solutions that fall off like powers of $1/r$ and so we expand all of the perturbations in a series in $1/r$, and solve the equations neglecting terms of order $1/r^2$ and higher. The resulting solutions have the following form
\be
 \delta \phi = \tilde H(r) + \hat H(r) \ln\(\frac{1-\cos\theta}{\sin\theta}\) -\frac12\hat A (r) \cos\theta \,,
\ee
\be
 \delta A = \tilde A(r) + \tilde C(r) -2 \widetilde W(r) + [ \hat C(r) -2 \,\widehat W(r) ] \ln\(\frac{1-\cos\theta}{\sin\theta}\) + \frac14 \hat A(r) \cos\theta \,,
\ee
\be
 \delta C = \tilde C(r) + \hat C(r) \ln\(\frac{1-\cos\theta}{\sin\theta}\) -\frac14\hat A(r) \cos\theta \,,
\ee
\be
 \delta B = \widetilde B(r) + \frac{ \tilde H(r)}{2} \, \theta \cot\theta - \hat H(r) \, \cM_2(\theta) - \hat A(r) \cos\theta\,,
\ee
\be
 \delta W= \widetilde W(r) + \widehat W(r) \, \ln\(\frac{1-\cos\theta}{\sin\theta} \) +\frac14\hat A(r) \, \cos\theta\,,
\ee
\be
 \frac{\delta F^{(0)}_{\rho\varphi}}{\ol F_{\rho\varphi}} = \widetilde B(r) + \tilde H(r) \(1+\frac\theta 2 \, \cot\theta\) + \hat H(r) \left[ \ln\(\frac{1-\cos\theta}{\sin\theta}\) -\cM_2(\theta) \right] - \frac52 \hat A(r) \, \cos\theta \,,
\ee
\be
 \hbox{and} \qquad
 \frac{\delta F_{r\varphi}}{\ol F_{\rho\varphi}} = - \frac{2}{r} \left( \frac{\hat C_0+\hat W_0}{\sin\theta} \right) \,.
\ee
where the function $\cM_2$ is defined by
\be
 \cM_2(x) := \int_0^x \exd y \; \frac{\cM_1(y)}{\sin^2 y} \quad
 \hbox{with} \quad  \cM_1(x) := \int_0^x\exd y \; \sin^2y\,\ln\(\frac{1-\cos y }{\sin y}\) \,,
\ee
as in Appendix \ref{Setup}.

In these expressions the integration constants\footnote{Our notation uses `tildes' for terms independent of $\theta$ and `hats' for constants that appear in $\theta$-dependent terms.} are $\tilde A_i$, $\hat A_i$, $\widetilde B_i$, $\tilde C_i$, $\hat C_i$, $\widetilde W_i$, $\widehat W_i$, $\tilde H_i$ and $\hat H_i$, where in all cases $i = 0,1$ and
\be
 \tilde A(r) := \tilde A_0 + \frac{\tilde A_1}{r} \,,
 \quad \hat A(r) := \hat A_0 + \frac{\hat A_1}{r} \,,
 \quad \hbox{and so on.}
\ee
The constants with subscript `0' appear in terms that survive as $r \to 0$ and describe those perturbations of the rugby ball that lead to the more general vacuum solutions given in \S\ref{sec:vacsolutions}. They can safely be set to zero for solutions describing point sources on a rugby ball solution. More generally their value is dictated by the near-brane matching conditions that set the vacuum solution that is appropriate between the branes far from the source.

\subsection{Boundary conditions}

We next eliminate the integration constants using the appropriate boundary conditions.

\subsubsection*{Near-brane boundary conditions}

We first impose the near-brane boundary conditions, eqs.~\pref{branematch1} and \pref{branematch2}. Of these, eqs.~\pref{branematch1} requires the near-brane limit, $\lim_{\theta \to 0} \theta \partial_\theta$, to vanish for the functions $\delta A$, $\delta C$, $\delta W$ and $\delta \phi$, and these conditions should hold for all $r$. They therefore require
\be
 \hat H(r) = \hat C(r) = \widehat W(r) = 0 \,.
\ee

Similarly, eq.~\pref{branematch2} fixes integration constants in $\delta B$ in terms of the brane lagrangians (which for the special case of rugby ball solutions are identical to one another). Physically, as shown in Appendix \ref{Setup}, the three quantities, $\widetilde B$, $\tilde H$ and $\hat A$ appearing in $\delta B$ can be interpreted geometrically as ($r$-dependent) modulations in the defect angles at each brane, and a change in the proper distance between the two branes. The near-brane matching condition requires these to be related for all $r$ to the constants appearing in the brane action in the same way as they are in the asymptotic, $r\to \infty$, geometry. In particular, if the background branes do not have an $r$-dependent action the asymptotic defect angles should not change, $\delta \alpha_\pm = 0$, which implies the parameters $\hat A$ and $\widetilde B$ must satisfy
\be
 \widetilde B = - \frac12 \, \tilde H \quad \hbox{and} \quad
 \hat A = 0 \,.
\ee
In particular, the condition $\hat A =0$ ensures that the functions $\delta A$, $\delta C$, $\delta W$ and $\delta \phi$ are all $\theta$-independent:
\be
 \delta \phi = \tilde H(r) \,, \quad
 \delta C = \tilde C(r) \,, \quad
 \delta W =\widetilde W(r) \,, \quad
 \delta A = \tilde A(r) + \tilde C(r) - 2 \widetilde W(r) \,,
\ee
\be
 \delta B =  \frac{ \tilde H(r)}{2} \Bigl( -1 + \theta \cot\theta \Bigr)
 \,, \quad
 \frac{\delta F_{\rho\varphi}}{\ol F_{\rho\varphi}} = \frac12 \, \tilde H(r) \Bigl( 1 + \theta  \, \cot\theta \Bigr)
 = \delta B + \delta \phi  \,,
\ee
and $\delta F_{r\varphi} = 0$. Furthermore, since having $\hat A=0$ negates the main advantage to our original definition of $\delta A$, we absorb the lingering $(\tilde C-2\widetilde W)$ into $\tilde A$ in what follows, so that $\delta A = \tilde A$ without loss of generality.

Now that we know that $\delta A$, $\delta C$ and $\delta W$ are independent of $\theta$, we know they are not independent of one another because it is possible to redefine $r$ in such a way as to impose one relation amongst them. If we restrict to transformations of the large-$r$ form $r \to r (1 + k/r + \cdots )$ only $\delta W$ can be changed in this way if terms of order $1/r^2$ are neglected, but this change can be used to set the constant $\widetilde W_1$ to any desired value.

\subsubsection*{Flux quantization}

We next impose the boundary condition coming from flux quantization, which states
\be
 \frac{1}{2\pi} \int \exd^2x \, F_{\theta\varphi} + \sum_{b = \pm} \frac{\gR^2 \cC_b e^{\phi_b}}{2\pi} = n \,,
\ee
where $n = \pm 1$ is the background quantum number, $\phi_b = \phi(\theta_b)$ and $\cC_b$ is the brane-localized flux coefficient of the brane lagrangian, eq.~\pref{eq:fluxq}. Linearizing and specializing to the case of a rugby ball asymptotic background (and so also to identical branes, $\cC_+ = \cC_-$) gives the condition
\be
 0 =  \cC\Bigl[ \delta \phi(\theta_+) + \delta \phi(\theta_-) \Bigr] + \frac{2\pi}{\gR^2}  \int_{\theta_+}^{\theta_-} \delta F_{\theta\varphi} = 2 \cC \tilde H \,,
\ee
from which we learn $\tilde H = 0$, leaving the solutions
\be \label{solnsCWA}
 \delta C = \tilde C(r) \,, \quad
 \delta W =\widetilde W(r) \,, \quad
 \delta A = \tilde A(r) \,,
\ee
and $\delta \phi = \delta B = \delta F_{\theta\varphi} = \delta F_{r\varphi} = 0$.

The function $\tilde H(r)$ physically describes an $r$-dependent modulation in the proper distance between the two branes, and it is flux quantization (together with the previous near-brane boundary conditions) that fixes this inter-brane distance for all $r$ in the same way that it did for the asymptotic vacuum solution at $r \to \infty$.

\subsubsection*{Far-field flux and source mass}

We finally examine boundary conditions appropriate to the large-$r$ limit, by trading the integration constant $\tilde C_1$ for the source mass, $M$, using eq.~\pref{eq:Mvslarger} or \pref{eq:Mvslarger1}, which in the present instance reduces to
\be \label{eq:Mvslargerbc}
 \tilde C_1 =  - \frac{\kappa^2 M}{32\pi^2 \alpha \, a_0^2} =  - \frac{\kappa_4^2 M}{8\pi}  = -G_4 M = - \frac{r_s}{2}  \,.
\ee

\subsection{The 4D Brans-Dicke perspective}

We now identify the physical interpretation of the remaining integration constants, by making contact with the low-energy 4D effective field theory. Before doing so we first digress to set aside a potential misconception about what the $1/r$ dependence of a field says about the existence of massless 4D degrees of freedom.

\subsubsection*{Light degrees of freedom, $1/r$ behaviour and integration constants}

At this point we naively have two related puzzles. First, we have used up our boundary conditions but are left with unspecified integration constants. Although $\tilde C_1$ can be traded for the source mass, $M$, and the constant $\widetilde W_1$ can be absorbed into a redefinition of $r$, the constant $\tilde A_1$ remains potentially unspecified. Second, these constants show that several independent metric components appear to fall off like $1/r$, potentially indicating the presence of a number of massless particles (besides just the graviton) in the 4D limit.\footnote{In this our calculation resembles that of ref.~\cite{PST}, who find for non-supersymmetric 6D systems that $1/r$ dependence in other fields besides the metric are important in reproducing the low-energy behaviour of 4D Einstein gravity.} We now argue that the existence of the extra integration constant is related to the existence of a light scalar degree of freedom, whose mass is nonzero but can be made parametrically light compared to the KK scale.

First a straw man: it is tempting to believe that if any 6D field behaves as $1/r$ at large distances (or, equivalently, approximately satisfies $\Box_4 \psi = 0$ at large distances), then it must describe an independent massless scalar field in the low-energy 4D effective theory. We now show why this is a fallacy, and that what is more important is the number of independent integration constants required to describe the long-distance behaviour of classical solutions.

To show why it is insufficient to know that a 6D field varies like $1/r$ to infer the existence of a massless mode, consider the following illustrative toy model of a heavy field, $h$, coupled to a light scalar field, $\ell$:
\be
 \cL = - \frac12 \Bigl[ (\partial h)^2 + (\partial \ell)^2 + M^2 \, h^2 + 2 \mu^2 \, h\, \ell + m^2 \ell^2 \Bigr] + J\, \ell \,.
\ee
Here we imagine $m \sim \mu \ll M$ and so $h$ and $\ell$ are almost mass eigenstates with the $h$ much heavier than $\ell$. The current $J$ represents the coupling to another sector ({\em e.g.} a point source), which we assume only couples to $\ell$.

Because this model is gaussian it can be solved exactly by diagonalizing the mass matrix, leading to propagation eigenstates
\be
 H = h \cos \vartheta + \ell \sin \vartheta \quad \hbox{and} \quad
 L = \ell \cos \vartheta - h \sin \vartheta \,,
\ee
where the mass eigenvalues are $M^2_\ssH \simeq M^2+\mu^4/M^2 \simeq M^2$ and $M^2_\ssL \simeq m^2 - \mu^4/M^2 \simeq m^2$ and the mixing angle is $\tan \vartheta \simeq \mu^2/M^2$.

Our interest is in the response of this system to a point source, and so in the solutions to the equations
\be
 ( -\Box + M_\ssH^2 ) H = J \sin \vartheta \quad \hbox{and} \quad
 ( - \Box + M_\ssL^2 ) L = J \cos \vartheta 
\ee
where $J = j \, \delta^{(3)} (x)$. For the purposes of the present parable we focus also on distances $M^{-1} \ll r \ll m^{-1}$ for which the mass of the field $L$ is not yet relevant, and it behaves as if it is effectively massless. In this regime we know the solution for $H$ falls exponentially to zero for $r \gg M^{-1}$, while $L$ varies as
\be \label{Lvsr}
 L \simeq \frac{j \, \cos \vartheta}{4\pi r} \simeq \frac{j}{4\pi r} \left( 1 - \frac{\mu^4}{2M^4} + \cdots \right) \,.
\ee
So far so good: the massive field falls off exponentially and the only effect of the mixing is to suppress the effective coupling of $L$ to the source.

Now consider instead analyzing this system by working with the initial basis $h$ and $\ell$ and perturbing in powers of $1/M$ from the get-go, as one does in practice when working with higher-dimensional theories. In this case the $h$ field equation is
\be
 (- \Box + M^2 )h + \mu^2 \ell = 0 \quad \hbox{and} \quad
  (- \Box + m^2)\ell + \mu^2 h  = J \,,
\ee
with approximate solution $h \simeq  -(\mu^2 / M^2) \ell + \cO \left( {1}/{M^4} \right)$. In the regime $r \gg m^{-1}, M/\mu^2$ the $h$ contribution to the $\ell$ equation is subdominant and so $\ell$ satisfies $ - \Box \ell \simeq J$, with approximate solution
\be \label{ellvsr}
 \ell \simeq \frac{j}{4 \pi r} \quad \hbox{and so} \quad
 h \simeq -\left( \frac{\mu^2 }{ M^2} \right) \frac{j}{4 \pi r} \,.
\ee

Notice that at this order $h$ acquires a $1/r$ profile (as appropriate given $\Box h \simeq -(\mu^2/M^2) \Box \ell \simeq 0$), even though the corresponding mass eigenstate $H$ is not at all an approximately light field. Of course eq.~\pref{Lvsr} shows why: virtual $h$ exchange must mediate a $1/r$ potential so that it can partially cancel the contribution of $\ell$ exchange, as given by eq.~\pref{ellvsr}, in order to reproduce the $\cos\vartheta$ dependence seen in \pref{Lvsr} when using proper mass eigenstates.

The lesson is this: having a heavy field vary as $1/r$ (or approximately satisfy $\Box h \simeq 0$) does not mean this is a new light field; instead it indicates that $h$ has a nonzero {\em overlap} with one of the light mass eigenstates.

A better way to count light degrees of freedom given only the $1/r$ dependence of a far-field solution is instead to count the independent integration constants. This is because each bona fide light degree of freedom appears as a field in the low-energy effective theory and as such has its own {\em independent} equation of motion in this theory. The extra integration constants arising when solving these equations represent the physical freedom to specify independently the initial conditions for this light field, separate from those of any other light fields in the problem. This is {\em not} what happened in the above toy example, where the solution $h \simeq -\mu^2 \ell/M^2$ dictates the integration constants of the $h$ field in terms of those of the $\ell$ field. This is a special case of the more general statement \cite{EFTrev} that the classical solutions of an effective field theory only capture the {\em adiabatic} solutions of the full theory involving heavier degrees of freedom.

In the present instance the `extra' integration constant is an indication that the low-energy far-field theory involves more than just the 4D graviton. In 6D supergravity the other light field is the 4D scalar `breathing mode', $\psi$, associated with 4D fluctuations of the parameter $\phi_0$ \cite{Linearized}. This parameter describes a flat direction of the equations of motion associated with the classical scaling symmetry described below eqs.~\pref{NSFEs}. Indeed, it would be exactly flat (at the classical level) in the absence of flux quantization, which lifts the flat direction by specifying $\phi_0$ through equations like eq.~\pref{fqrb}. Flux quantization can lift the flat direction \cite{LargeDim} because the brane-localized flux term (or $\cC$-dependent term) of the brane lagrangian, eq.~\pref{eq:fluxq}, breaks the scale invariance of the bulk field equations when $\cC$ is independent of $\phi$ (as is assumed here).

By breaking scale invariance the flux quantization condition gives the breathing mode a small mass, $\mu$, but this mass is parameterically small compared with the KK scale because it must `know' about the brane lagrangian, and so is suppressed relative to the KK scale by a power of $\delta = \kappa^2 L/2\pi \ll 1$. This suppression ensures there is a range of distance scales $\mu \ll 1/r \ll m_\KK \simeq a_0$ for which $r$ is large enough that a 4D description applies, but small enough that the mass $\mu$ is not yet relevant. Although the scalar should fall exponentially for $r \gg \mu^{-1}$, it falls approximately as $1/r$ within the regime between $a_0$ and $\mu^{-1}$. This is the regime for which the 6D $1/r$ solutions in general applies, and so the remaining integration constant should capture how the low-energy scalar modifies the system's response to a point source.\footnote{Of course, the 6D $1/r$ solutions also apply for $r \gg \mu^{-1}$, but in this regime the exponentially falling scalar profile instructs us to set the corresponding integration constant to zero.}

\subsubsection*{The 4D scalar-tensor effective field theory}

To fix the final integration constant we next match the linearized solution found above for the far-field 6D solution with the solution for an on-brane point source in the 4D EFT, assuming this EFT to be a scalar-tensor theory. The key for this matching is to recall that this solution is given in the 6D Einstein-frame metric
\be
g_{\ssM \ssN} \exd x^\ssM \exd x^\ssN = - e^{2 A} \exd t^2 + e^{-2 C} \exd r^2 + e^{2 W} r^2 \Bigl( \exd \xi^2 + \sin^2 \xi \, \exd \zeta^2 \Bigr) + a_0^2 \Bigl( \exd \theta^2 + e^{2 \ol B} \exd \varphi^2 \Bigr) \,,
\ee
with
\be \label{1rterms}
  A = \frac{\tilde A_1}{r} \,, \quad
  C = \frac{\tilde C_1}{r} \quad \hbox{and} \quad
  W = \frac{\widetilde W_1}{r}
\ee
as in eq.~\pref{solnsCWA}, and the dilaton and Maxwell field given by their background values, $\phi = \bar \phi$ and $F_{\ssM\ssN} = \ol F_{\ssM\ssN}$.

Dimensional reduction tells us that the 4D part of this metric is not in the 4D Einstein frame, but can be put there by making the rescaling $g_{\mu\nu} = (1/a_0)^2 \breve g_{\mu\nu}$, where $a_0$ is the radius of the extra dimensions in the asymptotic rugby-ball geometry. This is related to fluctuations in the canonically normalized breathing zero-mode, $\psi$, by $a_0 \propto e^{- \psi/2}$ and so $g_{\mu \nu} = e^{\psi} \, \breve g_{\mu\nu}$ \cite{SS, ABPQ}.

\subsection*{Matching to 4D Brans-Dicke theory}

The effective theory in the regime of interest for the above solution is the 4D effective theory of a light scalar coupled to the graviton \cite{sctensrev}, and the effective 4D lagrangian in this regime in the 4D Einstein frame is
\be
 \cL_\BD = - \frac{1}{2 \kappa_4^2} \, \sqrt{-\breve g} \; \breve g^{\mu\nu} \Bigl( \breve R_{\mu\nu} + \partial_\mu \psi \partial_\nu \psi +  \mu^2 \psi^2 \Bigr) + \cL_m \,,
\ee
where $\cL_m$ describes the coupling to other matter (such as the point source). As mentioned earlier, the nonzero mass, $\mu$, arises from flux quantization and is suppressed relative to the KK scale by factors of the defect angle: $\delta = \kappa^2 L/2\pi \ll 1$.

Our interest when matching to the 6D solutions is in regimes where the solutions fall off as $1/r$ and there are two separate such regimes: one where $a_0 \ll r \ll 1/\mu$ and both $\psi$ and the metric vary this way; and one with $r \gg 1/\mu$ for which $\psi$ vanishes exponentially and only the metric varies as $1/r$. To this end we neglect $\mu$, secure that we can reproduce the $r \gg \mu^{-1}$ regime simply by choosing the special case $\psi = 0$. Exterior to any sources the field equations obtained from this action therefore become
\be
 \breve R_{\mu\nu} + \partial_\mu \psi \partial_\nu \psi = 0
 \quad \hbox{and} \quad
 \breve \Box \, \psi = 0 \,,
\ee
for which the exact spherically symmetric solutions are \cite{STsolns}
\be
 \breve g_{\mu\nu} \exd x^\mu \exd x^\nu = - e^{2\breve A} \, \exd t^2 + e^{-2\breve C} \exd r^2 + r^2 e^{2\breve W} \, \Bigl( \exd \xi^2 + \sin^2 \xi \, \exd \zeta^2 \Bigr) \,,
\ee
where
\be
 e^{2\breve A} = e^{2\breve C} = \left( 1 - \frac{\ell}{r} \right)^p \,, \quad
 e^{2\breve W} = \left( 1 - \frac{\ell}{r} \right)^{1-p} \quad \hbox{and} \quad
 e^{\sqrt2 \; \psi} = \left( 1 - \frac{\ell}{r} \right)^q \,,
\ee
where the field equations impose one relation, $p^2 + q^2 = 1$, among the three integration constants, $p$, $q$ and $\ell$. These solutions are asymptotically flat, and in the far-field regime their linearization about the asymptotic flat geometry becomes
\be \label{deltaform4D}
 \delta \breve A = \delta \breve C = - \frac{p \,\ell}{2 r} \,, \quad
 \delta \breve W = - \frac{(1-p) \ell}{2r} \quad \hbox{and} \quad
 \delta \psi = - \frac{q \ell}{\sqrt2 \; r} \,.
\ee

To make contact with the 6D solution we write $g_{\mu\nu} = e^{\psi} \breve g_{\mu\nu}$, and so
\be
g_{\mu\nu} \exd x^\mu \exd x^\nu = - e^{2 A} \, \exd t^2 + e^{-2 C} \exd r^2 + r^2 e^{2 W} \, \Bigl( \exd \xi^2 + \sin^2 \xi \, \exd \zeta^2 \Bigr) \,,
\ee
with
\bea
 e^{2A} &=& e^\psi e^{2 \breve A} =  \left( 1 - \frac{\ell}{r} \right)^{p+q/\sqrt2} \simeq 1 - \frac{\left( p + q/\sqrt2 \right) \ell}{r} + \cdots \nn\\
 e^{2C} &=& e^{-\psi} e^{2 \breve C} =  \left( 1 - \frac{\ell}{r} \right)^{p-q/\sqrt2} \simeq 1 - \frac{\left( p - q/\sqrt2 \right) \ell}{r} + \cdots \nn\\
 \hbox{and} \quad  e^{2W} &=& e^\psi e^{2 \breve W} =  \left( 1 - \frac{\ell}{r} \right)^{1-p+q/\sqrt2} \simeq 1 - \frac{\left( (1-p) + q/\sqrt2 \right) \ell}{r} + \cdots  \,.
\eea
Comparing these expressions with eqs.~\pref{1rterms} then gives
\be
  A_1 = - \frac{\left( p + q/\sqrt2 \right) \ell}{2} \,, \quad
  C_1 = - \frac{\left( p - q/\sqrt2 \right) \ell}{2} \quad \hbox{and} \quad
  W_1 = - \frac{\left( 1-p + q/\sqrt2 \right) \ell}{2}  \,.
\ee
We ignore the $W_1$ equation, since this coefficient can be changed arbitrarily by redefining $r$.

\subsection*{Matching to sources in 4D}

We finally eliminate $p$, $q$ and $\ell$ in favour of the physical properties of the point source. If the point source on the brane in 6D does not couple directly to $\phi$ ({\em i.e.}~its microscopic mass is $\phi$-independent), then in 4D the source couples only to $\psi$ through a Brans-Dicke-like coupling \cite{BransDicke} to the 6D Einstein-frame (or `Jordan-frame') metric, $g_{\mu\nu} = e^{\psi} \breve g_{\mu\nu}$.

More generally, consider a generic matter action, $S_m$, that couples to $\psi$ and the metric only through the combination $g_{\mu\nu} := e^{2 \lambda(\psi)} \breve g_{\mu\nu}$, with $\lambda(\psi)$ a function that defines the specific theory. The choice $\lambda (\psi) = \lambda_0 \psi$ corresponds to the form described by Brans and Dicke themselves, and our 6D application is the further particular case\footnote{Notice that if $\psi$ had been massless a coupling $\lambda_0$ this large would be ruled out observationally by tests of general relativity within the solar system \cite{BDbounds}, although these bounds do not apply for the masses of practical interest here.} $\lambda_0 = \frac12$ \cite{ABS}. The equations of motion, including source terms, then become
\be
 \breve R_{\mu\nu} + \partial_\mu \psi \partial_\nu \psi + \kappa_4^2 \left( \breve T_{\mu\nu} - \frac12 \, \breve T \, \breve g_{\mu\nu} \right) = 0  \quad \hbox{and} \quad
  \breve \Box \, \psi + \kappa_4^2 \, \breve \cJ = 0 \,,
\ee
where
\be
\breve T^{\mu\nu} := \frac{2}{\sqrt{- \breve g}} \left( \frac{\delta S_m}{\delta \breve g_{\mu\nu}} \right) \quad \hbox{and} \quad
 \breve \cJ := \frac{1}{\sqrt{-\breve g}} \left( \frac{\delta S_m}{\delta \psi} \right)_{\breve g} = \lambda' \, \breve T \,.
\ee
Here the subscript `$\breve g$' indicates which metric is held fixed during the differentiation, while $\breve T := \breve g_{\mu\nu} \breve T^{\mu\nu}$ and $\lambda' := \exd \lambda/\exd \psi$.

Suppose now we have a non-relativistic source localized within a region $\Sigma$, for which only $\breve T_{tt}$ is significant and the gravitational and the source is weak enough that its gravitational binding energy is a negligible part of its rest mass. Then the mass is related to the asymptotic form of $\breve g_{tt}$ part of the metric by \cite{ADM}
\be
 \breve g_{tt} \simeq - 1 + \frac{r_s}{r} + \cO\left( \frac{r_s}{r} \right)^2 \,,
\ee
with $r_s = 2 G_4 M = \kappa_4^2 M/(4\pi)$. Comparing this with eqs.~\pref{deltaform4D} then gives $r_s =  p \,\ell$, from which we read
\be
 \ell = \frac{r_s}{p} = \frac{\kappa_4^2 M}{4\pi p} = \frac{2 G_4 M}{\sqrt{1-q^2}} \,.
\ee

Weak fields also imply
\be
 \psi \simeq \delta \psi = - \frac{q \ell}{\sqrt2 \; r} \,,
\ee
and so
\be
 \oint_r \exd S \; n \cdot \breve \nabla \psi \simeq 4\pi r^2 \left( \frac{\partial  \psi}{\partial r} \right) = \frac{4\pi q \ell}{\sqrt2} \,,
\ee
where the integration is over any sphere of radius $r$ surrounding, but exterior to, the source. On the other hand, when $\lambda' = \lambda_0$ is a constant integrating the dilaton field equation over space implies
\be
 \int \exd V \Bigl[\breve \Box \psi + \kappa_4^2 \breve \cJ \Bigr] = \oint_r \exd S \Bigl( n \cdot \breve \nabla \psi \Bigr) + \lambda_0 \kappa_4^2  \int \exd V \; \breve T
 = \frac{4\pi q \ell}{\sqrt2}  - \lambda_0 \kappa_4^2 M = 0 \,,
\ee
where we assume a non-relativistic source for which $\int \exd V \; \breve T \simeq - \int \exd V \; \breve T_{tt} \simeq M$. Solving for $q$ as a function of $M$ gives
\be
 \frac{q}{\sqrt2} = \frac{\kappa_4^2 \lambda_0 M}{4\pi\ell}  = \frac{\lambda_0 r_s}{\ell} = p \lambda_0 = \lambda_0 \sqrt{1-q^2} \,,
\ee
and so
\be
 \frac{q}{\sqrt2} = \frac{\lambda_0}{\sqrt{1 + 2\lambda_0^2}} \,, \quad
 p = \frac{1}{\sqrt{1 + 2\lambda_0^2}} \quad \hbox{and} \quad
 \ell = r_s \sqrt{1 + 2\lambda_0^2} \;,
\ee
(where we use $p \to 1$ as $\lambda_0 \to 0$).

This finally gives
\bea \label{BDmatching}
 \tilde A_1 &=&- \frac{\ell}{2} \left( p + \frac{q}{\sqrt2} \right) = -\frac{r_s}{2} \Bigl( 1 + \lambda_0 \Bigr) = - G_4 M \Bigl(1 + \lambda_0 \Bigr)\nn\\
 \hbox{and} \quad
 \tilde C_1 &=& - \frac{\ell}{2} \left( p - \frac{ q}{\sqrt2} \right) = - \frac{r_s}{2} \Bigl(1 - \lambda_0 \Bigr) = - G_4 M \Bigl(1 - \lambda_0 \Bigr) \,.
\eea
At first sight eqs.~\pref{BDmatching}  appear to contradict eq.~\pref{eq:Mvslargerbc}, which fixes $\tilde C_1$ purely in terms of the mass. In fact, they are compatible because eq.~\pref{eq:Mvslargerbc} really gives $\tilde C_1 = - G_4 m$, where $m$ is the {\em Jordan frame} mass, defined in terms of the Jordan-frame matter stress-energy:
\be
 T^{\mu\nu} := \frac{2}{\sqrt{- g}} \left( \frac{\delta S_m}{\delta g_{\mu\nu}} \right) = e^{-6\lambda} \breve T^{\mu\nu} \,.
\ee

Notice that using the value $\lambda_0 = \frac12$ appropriate for dimensional reduction gives
\be \label{BDmatchingn}
 \tilde A_1 = -\frac32 \, G_4 M \quad \hbox{and} \quad
 \tilde C_1 = -\frac12 \, G_4 M \qquad\qquad \hbox{if} \quad
 a_0 \ll r \ll \mu^{-1}\,.
\ee
Of course, for distances larger than the scalar mass, $r \gg 1/\mu$, the scalar field falls exponentially rather than like $1/r$, and so we instead effectively have $\lambda_0 = 0$ when seeking solutions as a power in $1/r$, leading to the usual 4D gravity result,
\be
 \tilde A_1 = \tilde C_1 = -G_4 M \qquad\qquad \hbox{if} \quad
 r \gg \mu^{-1}\,,
\ee
which, as expected, agrees with eq.~\pref{eq:Mvslargerbc} once the scalar plays no role.

\section{Gravitational phenomenology at small distances}
\label{sec:gpheno}

The discussion to this point points towards there being two interesting 4D regimes in the low-energy limit of the 6D supergravity.\footnote{The low-energy spectrum is even richer than this and includes other scalar fields lighter than the KK scale, such as those describing the brane positions. As described in more detail elsewhere \cite{4DSLED, branonbounds}, the gravitational effects of these other scalars are much smaller at practical distances than the Brans-Dicke scalar discussed here.} One enters the 4D regime at energies $E < m_\KK$ with a Brans-Dicke scalar-tensor gravity, and then passes over to 4D general relativity at energies below the 4D scalar mass, $E < \mu < m_\KK$. Furthermore, the expected Brans-Dicke coupling, $\lambda_0 = \frac12$, is not small and so scalar exchange can be important over distances shorter than $1/\mu$.

%\ignore
\FIGURE[h!]{
  \centering
	\includegraphics[width=0.45\textwidth]{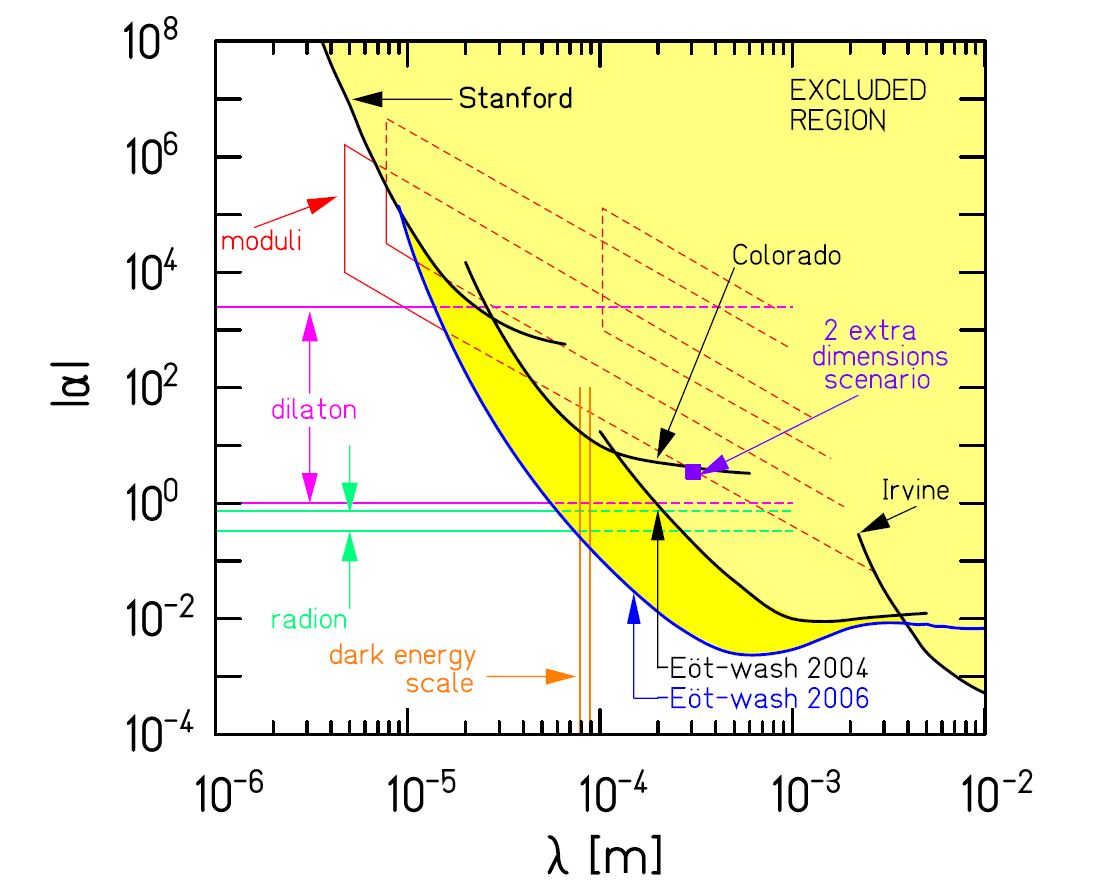}
    \caption{The 95\% confidence level exclusion plot for a Yukawa deviation from Newtonian gravity, in the $\alpha$--$1/\mu$ plane, as taken from ref.~\cite{InvSqTests}.}
\label{fig:Bounds}
}
%\endignore

This implies several potentially observable signals in short-distance gravitational tests. First, it means that deviations from Newton's inverse-square force law should start {\em before} these experiments get all the way down to the KK scale. Experiments testing the Newtonian prediction at short distances parameterize the deviations they seek in terms of a Yukawa potential of the form \cite{InvSqTests1}
\be
 V(r) = - \frac{G_4 M_1 M_2}{r} \Bigl( 1 + \alpha \, e^{- \mu \, r} \Bigr) \,,
\ee
and although this form is a fairly poor description of the crossover to the 6D potential $V \propto 1/r^3$ that occurs at the KK scale \cite{Callin}, it is precisely the potential --- with $\alpha = 2\lambda_0^2 = \frac12$ --- that is expected in the transition to the Brans-Dicke scalar. The present 95\% confidence exclusion plot, taken from ref.~\cite{InvSqTests}, is shown in Fig.~\ref{fig:Bounds}, from which we see that $\alpha < 1$ for $\mu^{-1} \simeq 45$ microns, strengthening to $\alpha \lsim 0.1$ for $\mu^{-1} \simeq 100$ microns. It also shows $\mu^{-1} \lsim 60$ $\mu$m for $\alpha = \frac12$. The upshot is that the normalization of the 4D Newton's constant should change by $\sim 50$\% as one passes below $r \sim \mu^{-1}$. Violations of the equivalence principle (such as a chemical-composition dependence to gravitational forces) would not be expected at these scales, however, because of the Brans-Dicke nature of the scalar coupling.

The key question then becomes: precisely how big is $\mu$ predicted to be? As argued above, because this mass vanishes in the absence of the branes, it is generically somewhat smaller than the KK scale, leading to the estimate
\be
   \mu \simeq m_\KK \, \delta \simeq \frac{\delta}{a_0}  \qquad \hbox{with} \quad \delta \simeq \frac{\kappa^2 T }{ 2\pi} \lsim 1 \,,
\ee
where $T$ is of order the brane tension. We must take $\delta$ smaller than unity to be sure to remain within the domain of validity of the semiclassical approximation, but it need not be enormously smaller.

It remains to estimate the allowed range for $m_\KK$ and $\delta$. The allowed range for these parameters is partly driven in these scenarios by the allowed range for the 6D gravity scale, $M_g = \kappa^{-1/2}$. A benchmark range of allowed values is $10 \; \hbox{TeV} \lsim M_g \lsim 40$ TeV, where the lower limit is required to evade astrophysical energy-loss bounds \cite{ADDastro, MSLED}, and the upper limit comes from requiring sufficiently small estimates \cite{DistSUSY} of the quantum contributions to the 4D vacuum energy. For this range the condition $M_p \propto M_g^2 \, a_0$ implies the KK scale lies in the sub-micron range $0.06 \; \mu\hbox{m} \lsim a_0 \lsim 1$ $\mu$m. A qualitative lower bound for the brane tension, $T$, comes from requiring it not to be smaller than $M_s^4$, where $M_s \simeq 5$ TeV is a conservative estimate for the lightest allowed excitations of Standard Model particles at the LHC. An upper limit on $T$ comes from the condition $\kappa^2 T = T/M_g^4 \ll 1$ required for control over the semiclassical approximation.

When $M_g = 10$ TeV we have $a_0 \simeq 1$ micron and so we see that $T \gsim (5 \; \hbox{TeV})^4$ requires $\delta \gsim \frac{1}{16}$ must be towards the upper end of its range. Taking the representative value $\delta \simeq 0.1$ gives $\mu^{-1} \simeq a_0/\delta \simeq 10$ microns. If, on the other hand, $M_g \simeq 40$ TeV then $a_0 \simeq 0.06$ $\mu$m and so taking $(5 \; \hbox{TeV})^4 \lsim T \lsim (20 \; \hbox{TeV})^4$ implies $0.01 \lsim \delta \lsim 0.1$ and so $0.6 \; \mu \hbox{m} \lsim \mu^{-1} \lsim 6 \; \mu \hbox{m}$.

Parameter ranges such as these would not yet be ruled out by tests of Newton`s laws, but are close to being experimentally accessible.

\section{Discussion}
\label{sec:discussion}

In summary, this paper constructs various exact and approximate solutions to the equations of 6D gauged, chiral supergravity that capture the response of the bulk supergravity fields to the presence of two space-filling branes and a point mass (that can be, but need not be) localized on one of these branes. Part of our goal in so doing is to have a 6D prediction to use when deriving the low-energy 4D effective field theory that is needed in order to describe efficiently cosmology in these models \cite{4DSLED}.

We argue that the near-horizon form of the solution for a small black hole resembles the corresponding solution for Einstein gravity in the bulk, because in the near-horizon limit the stress-energy associated with the other bulk fields becomes negligible relative to the curvature caused by the source.

We find a class of exact black-brane solutions that asymptote at large distances to the compactified bulk vacuum solutions, and argue that these also capture the far-field form of more general point-source configurations. In a nutshell, they do so because the multi-pole expansion guarantees that only the monopole moment is relevant in the far-field regime, and so in this regime any information about the position of the point source within the extra dimensions (that might distinguish the solution from a black brane) becomes lost.

We find the far-field solution has two physical integrations constants in it, after extra-dimensional near-brane and flux-quantization boundary conditions are imposed. We argue that these constants correspond to the mass and Brans-Dicke coupling that govern the same far-field solutions in the low-energy 4D effective theory that should apply in this limit. A scalar-tensor 4D theory is required because of the presence of the scalar breathing mode that is an exact flat direction of the classical equations of motion due to the classical rigid scale invariance of the 6D supergravity equations (a scale invariance that is common to most higher-dimensional supergravities).

This mode is known not to be exactly massless because the flux-quantization condition sees the brane-localized flux term of the brane action, which is the only part of the lagrangian that breaks the classical scale invariance. This lifting of the scale invariance by the branes allows flux quantization to determine the value of the flat-direction field, and thereby fix the size of the extra dimensions \cite{LargeDim, TNCC}. It also gives the corresponding 4D scalar a mass, and because this mass requires the presence of the brane (and its back-reaction) it is suppressed by a power of the brane defect angles, $\delta_\pm \simeq \kappa^2 T_\pm/2\pi$. As a consequence the scalar mass, $\mu$, is parametrically small compared with the KK scale, ensuring the existence of a range of scales described by a 4D scalar-tensor theory. For distances larger than $1/\mu$ the scalar can be integrated out, and the low-energy theory is just Einstein's gravity. For distances smaller than this, but larger than the KK scale, the appropriate 4D physics is gravity and a 4D scalar coupled to matter through a Brans-Dicke coupling.

In practice, as discussed in \cite{DistSUSY}, other considerations make it unlikely that the defect angle is fantastically small, because there is unlikely to be an enormous hierarchy between the  brane tension and higher-dimensional gravity scale. \S\ref{sec:gpheno} considered a benchmark range of parameters consistent with other constraints, and found that a low 6D gravity scale ($M_g \simeq 10$ TeV, which also produces the most favourable quantum contributions to the effective dark energy density \cite{DistSUSY}) would give an extra-dimensional size $a_0 \simeq 1$ micron and $\delta \simeq 0.1$, in which case $\mu^{-1} \simeq a_0/ \delta$ could be as large as $\mu^{-1} \simeq 10$ microns. If, on the other hand $M_g$, were as large as 40 TeV (which pushes up uncomfortably the quantum contributions to the vacuum energy) then $a_0 \simeq 0.06$ microns, and taking a plausible range of brane tensions leads to $\mu^{-1}$ between 0.6 and 6 microns. All estimates lead to observable changes in the micron range, and the smaller $\delta$ is the larger the intermediate range of scales for which a 4D Brans-Dicke description applies. 

This indicates the likelihood of an interesting rich gravitational phenomenology in the 1--10 micron regime. The inverse-square law should persevere down to $m_\KK^{-1} \simeq 1$ micron, but the strength of Newton's constant should change by an order-unity amount as the scalar-tensor nature of gravity becomes manifest at $\mu^{-1} \simeq 10$ microns. There should be no violations of the equivalence principle at 10 microns due to the Brans-Dicke nature of the scalar-matter couplings. Eventually the crossover to the 6D $1/r^4$ power-law for the force occurs at the KK scale itself.

In the long run, predictions such as these should be hard to miss, particularly once joined with the other observable predictions at KK scales \cite{Callin} and for particle physics \cite{MSLED, ADDpheno, SLEDpheno, stringLHC} made by the SLED picture. In a subsequent publication \cite{4DSLED} we hope to explore in more detail the implications of the low-energy limit found here, both for laboratory experiments and for cosmology, by scoping out the 4D effective theory in more detail.

It would be remarkable if the puzzle of the vacuum energy --- normally regarded as the most depressing of problems, devoid of progress despite much study --- were to point towards such a rich diversity of new phenomena right on our intellectual doorstep.

\section*{Acknowledgements}

We thank Niayesh Afshordi, Kurt Hinterbichler, Hyun-Min Lee, Susha Parameswaran, Marco Peloso, Maxim Pospelov, Alberto Salvio, Gianmassimo Tasinato and Itay Yavin for useful discussions. We thank the Abdus Salam International Centre for Theoretical Physics (ICTP) for its support and the pleasant environs within which part of this work was done. Our research was supported in part by funds from the Natural Sciences and Engineering Research Council (NSERC) of Canada. Research at the Perimeter Institute is supported in part by the Government of Canada through Industry Canada, and by the Province of Ontario through the Ministry of Research and Information (MRI).

\appendix

\section{Newtonian limit}
\label{app:newtonian}

In this limit we consider the Newtonian limit of a point-mass source, for which the metric is $g_{\ssM \ssN} = \eta_{\ssM \ssN} + h_{\ssM \ssN}$ and $T_{\ssM \ssN} = M \, \delta^{n-1}(x) \, \delta_\ssM^t \delta_\ssN^t$, where $n = 4$ or 6 is the dimension of spacetime. Our goal is to derive the Newtonian expressions for $r_s$ and $\rho_s$ in terms of the mass $M$.

It is most convenient to use the trace-reversed field equations, which in $n$ dimensions is
\be
 \cR_{\ssM \ssN} + \kappa_n^2 \left( T_{\ssM \ssN} - \frac{T}{n-2} \, g_{\ssM \ssN} \right) = 0 \,,
\ee
where $T := g^{\ssM \ssN} T_{\ssM \ssN}$. In this we use the de Donder gauge result for the Ricci tensor, $\cR_{\ssM \ssN} = \frac12 \, \Box \, h_{\ssM \ssN}$ (using Weinberg's curvature conventions). Since only $T_{tt}$ is nonzero $T = g^{tt} T_{tt}$. This leaves the $(tt)$ Einstein equation
\be
  \Box \, h_{tt} + 2\left( \frac{n-3}{n-2} \right) \kappa_n^2 T_{tt} \simeq 0 \,.
\ee

\subsubsection*{4 dimensions}

Specializing to 4D and $T_{tt} = M \, \delta^3(x)$ we have $n=4$ and so
\be
 - \Box \, h_{tt} = \kappa_4^2 M \, \delta^3(x) \,,
\ee
whose solution is
\be
 h_{tt} = \frac{\kappa_4^2 M}{4 \pi \, r} \,.
\ee
Using $g_{tt} \simeq - \Bigl( 1 + 2 \Phi \Bigr)$, where $\Phi$ is the Newtonian potential, this becomes
\be
 \Phi = - \frac{\kappa_4^2 M}{8 \pi \, r} = - \frac{G_4 M}{r} \,,
\ee
where $G_4$ is Newton's constant and we use $\kappa_4^2 = 8 \pi G_4$. (Indeed, this is how this relation between $\kappa_4$ and $G_4$ is derived.) In terms of this we have
\be
 g_{tt} = \eta_{tt} + h_{tt} = -1 - 2 \Phi = - \left( 1 - \frac{r_s}{r} \right)
\ee
where
\be
 r_s = \frac{\kappa_4^2 M}{4 \pi} = 2 G_4 M \,,
\ee
as usual.

\subsubsection*{6 dimensions}

Next specializing to 6D and $T_{tt} = m \, \delta^5(x)$ we have $n=6$ and so
\be
 -\Box \, h_{tt} = \frac{3\kappa_6^2 M }2 \, \delta^5(x) \,,
\ee
whose solution is
\be
 h_{tt} =  \left( \frac{3\kappa_6^2 M }{2} \right) \frac{1}{8 \pi^2  \rho^3} = \frac{3 \kappa_6^2 M}{16 \pi^2  \rho^3} \,,
\ee
where we use $\rho$ to denote the 6D radial coordinate.

Using $g_{tt} \simeq - (1 + 2 \Phi)$, where $\Phi$ is the Newtonian potential, this becomes
\be
 \Phi = - \frac{3\kappa_6^2 M}{32 \pi^2 \, \rho^3} = - \frac{3G_6 M}{4\pi \rho^3} \,,
\ee
if we define $G_6$ in terms of $\kappa_6$ through the same relation as in 4D, $\kappa_6^2 = 8 \pi G_6$.

In terms of this we have
\be
 g_{tt} = \eta_{tt} + h_{tt} = -1 - 2 \Phi = - \left( 1 - \frac{\rho_s^3}{\rho^3} \right)
\ee
where
\be \label{rhos}
 \rho_s^3 = \frac{3 \kappa_6^2 M}{16 \pi^2} \,.
\ee

\section{Conserved currents and surface integrals}
\label{app:conservedcurrents}

This Appendix derives the expressions for the conserved charges as surface integrals, as is used in the main text to find $r_s$ and $\rho_s$ as functions of $M$ in the fully relativistic setting.

The equations of motion used in the main text are
\begin{gather} \label{app:sugraFE}
 {\cE^\ssM}_\ssN := {\cG^\ssM}_\ssN + \kappa^2 {T^\ssM}_\ssN = 0 \quad{\rm where}\quad\\
 \kappa^2 {T^\ssM}_\ssN := \pd^\ssM \phi \,\pd_\ssN \phi - \frac12 \, (\pd \phi)^2\,{\delta^\ssM}_\ssN + \frac{\kappa^2 e^{-\phi}}{\gR^2} \left( F^{\ssM\ssP} F_{\ssN\ssP} - \frac14 \, F^2 \, {\delta^\ssM}_\ssN \right) - \frac{2\gR^2}{\kappa^2} \, e^\phi \, {\delta^\ssM}_\ssN  + {t^\ssM}_\ssN\,, \nn\\
\pd_\ssM \Big(\sqrt{-g} e^{-\phi} F^{\ssM\ssN} \Big) + j^\ssN = 0 \,, \\
\square \phi + \left( \frac{\kappa^2e^{-\phi}}{4g^2} F_{\ssP\ssQ} F^{\ssP\ssQ} - \frac{2g^2}{\kappa^2} e^\phi \right) = 0 \,,
\end{gather}
where $t_{\ssM\ssN}$ and $j^\ssN$ are the brane and particle stress-energy and gauge current.

\subsection*{Conserved charges}

Given a metric, $g_{\ssM \ssN}$, that asymptotes to $\bar g_{\ssM \ssN}$ at infinity, we can define a conserved charge $Q$ for each isometry of $\bar g_{\ssM \ssN}$. Given a Killing vector field (KVF), $V^\ssM$, of $\bar g_{\ssM \ssN}$,
\be \label{KVF}
 \ol \nabla_\ssM \ol V_\ssN + \ol \nabla_\ssN \ol V_\ssM = 0 \,,
\ee
where $\ol V_\ssM := \bar g_{\ssM \ssN} V^\ssN$, we define the background-covariantly conserved (pseudo)current,
\be
 \cJ^\ssM := - \frac{1}{\kappa^2} \, {\Delta \cE^\ssM}_\ssN V^\ssM \,,
\ee
where $\Delta {\cE^\ssM}_\ssN$ contains all terms of the field equation, ${\cE^\ssM}_\ssN$, that are nonlinear in the difference $\delta g_{\ssM\ssN} := g_{\ssM \ssN} - \bar g_{\ssM \ssN}$ (and similarly for the deviations of any other fields).

In terms of this the charge is defined by
\be
 Q := \int_\Sigma \exd \Sigma_\ssM \cJ^\ssM = - \frac{1}{\kappa^2} \int_\Sigma \exd^5x \sqrt{\bar g} \; n_\ssM {\Delta \cE^\ssM}_\ssN V^\ssN \,,
\ee
where $\Sigma$ is a time-like hypersurface with unit normal $n_\ssM$. In practice we work with a diagonal metric and take $\Sigma$ to be surfaces of constant $t$, so $n_\ssM \exd x^\ssM = \sqrt{-g_{tt}} \; \exd t$. This charge is conserved in the sense that it is independent of $\Sigma$, and this can be seen given that the Bianchi identity, $\ol \nabla_\ssM {\Delta \cE^\ssM}_\ssN = 0$, with the KVF condition, eq.~\pref{KVF}, implies $\ol \nabla_\ssM J^\ssM = 0$.

The connection to asymptotic fields is made by using the field equations, ${\cE^\ssM}_\ssN = 0$ satisfied by $g_{\ssM \ssN}$, and ${\ol\cE^\ssM}_\ssN = 0$ satisfied by $\bar g_{\ssM \ssN}$, since these imply ${\delta \cE^\ssM}_\ssN + \Delta {\cE^\ssM}_\ssN = 0$, where ${\delta \cE^\ssM}_\ssN$ is the part of the field equations linear in $\delta g_{\ssM\ssN}$ (and other perturbations). These equations imply that the conserved charge may be written
\be
 Q = \frac{1}{\kappa^2} \int_\Sigma \exd \Sigma_\ssM {\delta \cE^\ssM}_\ssN V^\ssN \,.
\ee
What is useful about this is that general covariance implies the integrand is always a total derivative, and so ${\delta \cE^\ssM}_\ssN V^\ssN = \ol \nabla_\ssL s^{\ssL \ssM}$ for some $s^{\ssL \ssM}$. As a consequence $Q$ can be written as a surface integral over the boundary, $\partial \Sigma$, of $\Sigma$:
\be
 Q = \frac{1}{\kappa^2} \oint_{\partial\Sigma} \exd \Sigma_{\ssM\ssN} s^{\ssM\ssN} \,.
\ee

In particular the total mass, $M$, of a configuration can be defined as the conserved charge associated with the time-like KVF, $V^\ssM \partial_\ssM = \partial_t$, for which
\be
 M = \frac{1}{\kappa^2} \int_{\Sigma} \exd^5x \sqrt{- \bar g} \; {\delta \cE^t}_t \,.
\ee
To see how the conserved charges work in detail it is worth working through several explicit examples.

\subsubsection*{Spherical 6D black hole}

In this case we use Einstein's equations, $\cG_{\ssM \ssN} + \kappa^2 T_{\ssM\ssN} = 0$, where the metric is asymptotically flat, so $\bar g_{\ssM \ssN} = \eta_{\ssM \ssN}$. We take a point mass as the source (or can consider it to be a black hole).

The metric in this case is
\be
 \exd s^2 = - h(\rho) \, \exd t^2 + \frac{\exd \rho^2}{h(\rho)} + \rho^2 \exd^2 \Omega_4 \,,
\ee
where $\exd^2 \Omega_4$ is the metric on the unit 4-sphere (whose volume is $\cV_4 = \frac83 \, \pi^2$). For a Schwarzschild metric the function $h(\rho)$ is given by
\be \label{hform}
 h(\rho) = 1 - \frac{\rho_s^3}{\rho^3} \,,
\ee
where $\rho_s$ is a length to be determined in terms of $M$ (and compared with eq.~\pref{rhos}, above).

As above, we define the mass within a region of radius $\rho$ by
\be
 M = - \frac{1}{\kappa^2} \int_\Sigma \exd^5 x \sqrt{- \bar g} \; {\Delta \cE^t}_t \,.
\ee
In general $M$ defined in this way includes the gravitational binding energy, to the extent that the nonlinear terms in the Einstein tensor contribute to $\Delta {\cE^t}_t$. But things are simpler when the fields are weak enough that it suffices to work to linear order in ${\cG^t}_t$, since in this case the only contribution comes from the stress energy of the source, leading to
\be
 M =- \frac{1}{\kappa^2} \int_\Sigma \exd^5 x \sqrt{- \bar g} \; {\Delta \cE^t}_t = - \int_\Sigma \exd^5 x \sqrt{-\bar g} \; {T^t}_t \,.
\ee

Now consider evaluating the linearized part of the Einstein equation. For pure gravity this is given by
\be
 {\delta \cE^t}_t = \eta^{tt} \delta \cG_{tt} = - \frac{2}{\rho^2} \Bigl( 3 \, \delta h + \rho \, \partial_\rho \delta h \Bigr) \,,
\ee
and so
\be
 M = \frac{1}{\kappa^2} \int_\Sigma \exd^5 x \sqrt{- \bar g} \; {\delta \cE^t}_t = \frac{8 \pi^2 \alpha}{3 \kappa^2} \int_0^{\rho_\star} \exd \rho \, \rho^4 {\delta \cE^t}_t  = - \frac{16 \pi^2 \alpha}{3 \kappa^2} \int_0^{\rho_\star} \exd \rho \, \partial_\rho \Bigl( \rho^3 \, \delta h \Bigr)  \,,
\ee
where $\rho_\star$ is a particular radius outside of the source. We include a defect angle $\delta = 2\pi(1 - \alpha)$, to include the case of a brane threading the source. Using eq.~\pref{hform} to infer $\delta h = - \rho_s^3/\rho^3$ allows the result to be evaluated, and gives
\be \label{SchMvsrhos}
 M = - \frac{16 \pi^2 \alpha}{3 \kappa^2} \Bigl[ \rho_\star^3 \, \delta h(\rho_\star) \Bigr] = \frac{16 \pi^2 \alpha \, \rho_s^3}{3 \kappa^2} \,.
\ee
Solving this for $\rho_s$ gives
\be
 \rho_s^3 = \frac{3 \kappa^2 M}{16 \pi^2 \alpha} \,,
\ee
in agreement with eq.~\pref{rhos} (and generalizing it to $\alpha \ne 1$).

\subsection*{Applications in 6D supergravity}

Our real system of interest is the supergravity of eqs.~\pref{app:sugraFE}, for which we now repeat the same construction. In this case we take a diagonal metric {\it ansatz},
\bea \label{genansatz}
 \exd s^2 &=& \exd s_4^2 + a_0^2 \big(e^{2E(\theta,r)} \exd \theta^2 + e^{2B(\theta,r)} \exd\varphi^2 \big)  \\
 \exd s_4^2 &=& - e^{2A(\theta,r)} \, \exd t^2 + e^{-2C(\theta,r)} \, \exd r^2+ e^{2W(\theta,r)}  r^2 (\exd \xi^2 + \sin^2\xi \, \exd \zeta^2 ) \,,
\eea
which we linearize about the asymptotic rugby-ball solution, for which $\ol A = \ol C = \ol W = \ol E = 0$ and where
\be
 e^{\ol B} = \alpha \sin\theta \,,\quad \ol F_{\theta\varphi} = \pm\frac{\alpha\sin\theta}{2}  \,,\quad e^{\ol\phi} = e^{\phi_0} := \frac{\kappa^2}{4\gR^2a_0^2}  \,.
\ee

In this case the $(tt)$ components of the Einstein tensor and the bulk stress-energy tensor become
\bea
 {\ol \cG^t}_t &=& \frac1{a_0^2} + \frac{1}{a_0^2} \, \pd_\theta^2 ( \delta C - \delta B - 2 \delta W) + \frac{\cot\theta}{a_0^2} \pd_\theta( \delta C + \delta E - 2 \delta B - 2 \delta W) \\
 &&\quad - \pd_r^2( \delta E + \delta B + 2 \delta W) - \frac2r \pd_r (\delta C + \delta E + \delta B + 3 \delta W) - \frac{2 \delta E}{a_0^2} - \frac{2(\delta C + \delta W)}{r^2} \nn\\
 \kappa^2 {\ol T^t}_t &=& \frac1{a_0^2} \left( -1 + \delta E + \delta B - \frac{\delta F_{\theta\varphi}}{\ol F_{\theta\varphi}} \right) \,,
\eea
and so, after some rearranging, we find that these combine to give the following total derivative for ${\delta \cE^t}_t$:
\bea
 \sqrt{- \bar g} \; {\delta \cE^t}_t &=& \sqrt{- \bar g} \, \big({\delta \cG^t}_t + \kappa^2 {\delta T^t}_t\big) \nn\\
 &=& \frac{1}{a_0^2} \, \pd_\theta \left\{ \sqrt{-\bar g} \left[ \pd_\theta \Bigl( \delta C - \delta B - 2 \delta W \Bigr) + \Bigl(\delta E - \delta B \Bigr) \cot\theta  - \frac{\delta A_\varphi}{\ol F_{\theta\varphi}}\right] \right\} \nn\\
&&\quad -\pd_r \left\{ \sqrt{-\bar g} \left[ \pd_r \Bigl( \delta E + \delta B + 2 \delta W \Bigr) + \frac{2}r \, \Bigl(\delta C + \delta W \Bigr) \right] \right\} \,.
\eea

In terms of this, the conserved mass is obtained by integrating. Assuming all of the functions are functions only of $r$ and $\theta$ also allows some of the remaining angular integrations to be performed explicitly, since integrating $\xi$ and $\zeta$ gives $4\pi$ and the integral over $\varphi$ gives $2\pi$. Using $\sqrt{- \bar g} = \alpha \,a_0^2\, r^2 \sin \xi \sin \theta$ and integrating over a region $0 < r < r_\star$ and $\theta_1 < \theta < \theta_2$ gives the following result for $M$
\bea \label{Msugra}
 M &=& \frac{1}{\kappa^2} \int_{\Sigma(r_\star,\theta_i)} \exd^5x \sqrt{-\bar g} \; {\delta \cE^t}_t \nn\\
 &=& \frac{8\pi^2 \alpha}{\kappa^2} \int_0^{r_\star} \exd r \left\{ r^2 \sin \theta \left[ \pd_\theta \Bigl( \delta C - \delta B - 2 \delta W \Bigr) + \Bigl(\delta E - \delta B \Bigr) \cot\theta  - \frac{\delta A_\varphi}{\ol F_{\theta\varphi}}\right] \right\}_{\theta=\theta_1}^{\theta=\theta_2} \nn\\
 && \;\; - \frac{8\pi^2 \alpha \, a_0^2}{\kappa^2} \int_{\theta_1}^{\theta_2} \exd \theta \left\{ r^2 \sin \theta \left[ \pd_r \Bigl( \delta E + \delta B + 2 \delta W \Bigr) + \frac{2}{r} \, \Bigl(\delta C + \delta W \Bigr) \right] \right\}_{r=0}^{r=r_\star}  \,.
\eea

\subsubsection*{The 6D Schwarzschild black hole (again)}

As a first application (and a reality check) we can apply this formula to compute the mass of a 6D Schwarzschild black hole, which should provide an approximate solution to these equations over distances much smaller than the KK scale, $a_0$. To this end it is worth rewriting the Schwarzschild metric in the following alternative form,
\be \label{appeq:6DBHcyl1}
 \exd s^2 = - \left( \frac{4 \varrho^3 - \rho_s^3}{4 \varrho^3 + \rho_s^3} \right)^2
 \exd t^2 + \left( 1 +  \frac{\rho_s^3}{4\varrho^3} \right)^{4/3} \exd s_5^2 \,,
\ee
where
\be \label{appeq:6DBHcyl2}
 \exd s_5^2 = \exd r^2 + r^2 \Bigl( \exd \xi^2 + \sin^2 \xi \, \exd \zeta^2 \Bigr) + a_0^2 \Bigl( \exd \theta^2 + \alpha^2 \theta^2 \, \exd \varphi^2 \Bigr) \,,
\ee
is the flat metric on a 5D cone with $\alpha = 1 - \delta/2\pi$ measuring the defect angle. Here $\varrho$ is related to the other coordinates by $\varrho^2(r, \theta) = r^2 + a_0^2 \theta^2$, and to the 6D Schwarzschild radial coordinate by
\be
 \rho = \varrho \left( 1 + \frac{\rho_s^3}{4 \varrho^3} \right)^{2/3}
 \simeq \varrho + \frac{\rho_s^3}{6 \varrho^2} + \cdots \,,
\ee
where the last, approximate equality assumes both $\rho$ and $\varrho$ to be much larger than $\rho_s$.

This should be a solution to the supergravity equations, eqs.~\pref{app:sugraFE}, over distances much smaller than the KK size, and so for which we may use $\sin \theta \simeq \theta$ within the extra-dimensional metric. In this regime this form of the black-hole metric agrees with the general ansatz of eq.~\pref{genansatz}, and a comparison shows that the metric functions are
\be
 e^{2A} = \left( \frac{4 \varrho^3 - \rho_s^3}{4 \varrho^3 + \rho_s^3} \right)^2 \quad \hbox{and} \quad
 e^{-2 C} = e^{2W} = e^{2 \delta B} = e^{2E} = \left( 1 +  \frac{\rho_s^3}{4\varrho^3} \right)^{4/3} \,,
\ee
and so the linearized deviations from the rugby ball are
\be
 \delta A = - \frac{\rho_s^3}{2\,\rho^3} \qquad \hbox{and} \qquad
 -\delta C = \delta W = \delta E = \delta B = \frac{\rho_s^3}{6 \rho^3} \,.
\ee

Specializing eq.~\pref{Msugra} to the case $-\delta C = \delta E = \delta B = \delta W$ and $\delta A_\varphi = 0$, and choosing $\theta_1 = 0$ and $\theta_2 = \theta_\star$, gives
\be \label{MsugraBH}
 M = -\frac{32\pi^2 \alpha}{\kappa^2} \left\{ \int_0^{r_\star} \exd r \Bigl[ r^2 \sin \theta \, \pd_\theta \delta W  \Bigr]_{\theta=0}^{\theta=\theta_\star} + a_0^2 \int_{0}^{\theta_\star} \exd \theta \Bigl[ r^2 \sin \theta \, \pd_r  \delta W  \Bigr]_{r=0}^{r=r_\star} \right\}\,.
\ee
In this use the explicit form for $\delta W$ and perform the derivatives using $\rho^2 = r^2 + u^2$ where $u := a_0 \theta$,
\be
 \partial_\theta \delta W = a_0 \partial_u \delta W =  - \frac{a_0 \rho_s^3 u}{2 \rho^5} \quad \hbox{and} \quad
 \partial_r \delta W = - \frac{\rho_s^3 r}{2 \rho^5} \,.
\ee
Furthermore, we also take $\theta_\star$ small enough that $a_0 \sin \theta \simeq a_0 \theta = u$, since this is the regime in which we expect the Schwarzschild solution approximates a solution to eqs.~\pref{app:sugraFE}. The result is
\bea \label{SchMsugra}
 M &=& \frac{16\pi^2 \alpha \, \rho_s^3}{\kappa^2} \left\{ \left[ u^2 \int_0^{r_\star}  \frac{r^2 \, \exd r }{\rho^5}  \right]_{u=0}^{u=a_0\theta_\star} + \left[ r^3 \int_{0}^{a_0 \theta_\star}  \frac{u \, \exd u }{\rho^5}  \right]_{r=0}^{r=r_\star} \right\} \nn\\
 &=& \frac{16\pi^2 \alpha \, \rho_s^3}{\kappa^2} \left[ \left( \frac{r_\star^3}{3\rho_\star^3} \right) + \frac13 \left( 1 - \frac{r_\star^3}{\rho_\star^3} \right) \right] \nn\\
 &=& \frac{16\pi^2 \alpha \, \rho_s^3}{3\kappa^2} \,,
\eea
which defines $\rho_\star^2 := r_\star^2 + u_\star^2 = r_\star^2 + a_0^2 \theta_\star^2$ and uses the integrals
\be
 \int  \frac{r^2 \, \exd r }{(r^2 + u^2)^{5/2}} = \frac{r^3}{3 u^2 (r^2+u^2)^{3/2}} \quad
 \hbox{and} \quad
  \int  \frac{u \, \exd u }{(r^2 + u^2)^{5/2}} = \frac{1}{3 (r^2 + u^2)^{3/2}} \,.
\ee
Notice that eq.~\pref{SchMsugra} agrees with the result, eq.~\pref{SchMvsrhos}, obtained using the Schwarzschild solution and the conserved charge for Einstein gravity in 6D spherical coordinates.

\section{Far-field supergravity solutions}
\label{Setup}

This appendix describes in more detail the linearization of the bulk field equations about the rugby-ball solution, and their integration to obtain the far-field solution given in the main text. We follow closely the approach of ref.~\cite{LargeDim}.

We start with the metric {\it ansatz}
\be
 \exd s^2 = - e^{2A(\theta,r)} \, \exd t^2 + e^{-2C(\theta,r)} \, \exd r^2+e^{2W(\theta,r)}\, r^2 (\exd \xi^2 + \sin^2\xi \, \exd \zeta^2 )  + a_0^2\big(\exd \theta^2 + e^{2B(\theta,r)} \,\exd\varphi^2\big) \,,
\ee
and also allow $\phi$ and the gauge fields to depend on $r$ and $\theta$.
The non-vanishing components of the 6D Ricci tensor for this metric are
\bea
 -e^{-2A}\,\cR_{tt} &=& e^{2C} \bigg[\pd_r^2 A + \pd_r A \bigg( \frac2r + \pd_r(A+ B+C+2W) \bigg)\bigg] \nn\\
 && \qquad
 + \frac1{a_0^2}\Big(\pd_\theta^2 A  + \pd_\theta A \, \pd_\theta(A-C+2W+B) \Big) \qquad  \nn\\
 e^{2C}\,\cR_{rr} &=& e^{2C} \bigg[\pd_r^2 A + (\pd_r A)^2 + \pd_r C \bigg( \frac2r + \pd_r (A+B+2W) \bigg) + \pd_r^2 B + (\pd_r B)^2 \nn\\
 && \qquad
 +2 \,\pd_r^2 W + 2(\pd_r W)^2 \bigg]- \frac1{a_0^2}\Big(\pd_\theta^2 C  + \pd_\theta C \, \pd_\theta(A-C+2W+B)\Big) \nn\\
 e^{-2W}\,\cR_{\xi\xi} &=& -e^{-2W} + e^{2C} \Big[1+r^2 \,\pd_r^2 W + r^2 \,\pd_r W \,\pd_r (A+B+C+2W) \\
 && \qquad 
  + r \,\pd_r (A+B+C+4W) \Big] + \frac{r^2}{a_0^2} \Big[ \pd_\theta^2 W + \pd_\theta W\, \pd_\theta \big( A-C+2W+B\big)\Big] \nn\\
 \cR_{\zeta\zeta} &=& \cR_{\xi\xi} \sin^2\xi \nn\\
 \cR_{\theta\theta} &=& \pd_\theta^2 (A+B-C+2W) + (\pd_\theta A)^2+ (\pd_\theta B)^2 + (\pd_\theta C)^2 + 2(\pd_\theta W)^2 \nn\\
 \cR_{\theta r} &=& \pd_\theta \pd_r (A+B+2W) + \pd_\theta A \, \pd_r A + \pd_\theta B \, \pd_r B +2\,\pd_\theta W\,\pd_r W  \nn\\
 && \qquad
 +\pd_\theta C \, \pd_r (A +B+2W) + \frac2r (\pd_\theta C +\pd_\theta W)\qquad \nn\\
 e^{-2B} \,\cR_{\varphi\varphi} &=& a_0^2 e^{2C} \bigg[ \pd_r^2 B +\pd_r B \bigg(\frac2r + \pd_r( A+B+C+2W) \bigg)\bigg] \nn\\
 && \qquad
 + \pd_\theta^2 B + \pd_\theta B \,\pd_\theta(A-C+2W+B) .\qquad \nn
\eea

\subsection*{The linearized equations}

We wish to linearize the field equations about the rugby-ball solution, which corresponds to the choices
\be
 \ol A = \ol C = \ol W = 0 \quad \hbox{and} \quad
 e^{\ol B} =  \alpha \, \sin \theta \,,
\ee
with
\be
 \ol \phi = \phi_0 = \ln\left(\frac{\kappa^2}{4g^2a_0^2}\right)
\ee
and
\be
 \ol F_{\theta\varphi} = \frac{\alpha \, n}{2} \, \sin\theta \,,
\ee
where $n=\pm1$. To linearize we write
\bea
 && A = \ol A + \delta A \,, \quad B = \ol B + \delta B \,,
 \quad C = \ol C + \delta C \,,\nn\\
 &&W = \ol W + \delta W \,, \quad \phi = \ol \phi + \delta \phi \,,\quad F_{\ssM\ssN} = \ol F_{\ssM\ssN}  + \delta F_{\ssM\ssN} \,,
\eea
in the field equations and drop all quadratic terms, with the goal of finding the general solution for the resulting linear differential equations for the variations.

The linearized Ricci-tensor components become
\bea
 \cR_{tt} &=& -\ol\Box \delta A \nn\\
 \cR_{rr} &=& - \ol\Box_2 \delta C + \pd_r^2 (\delta A+\delta B+2\,\delta W) + \frac2r \, \pd_r(\delta  C+2\,\delta W) \nn\\
 \cR_{\xi\xi} &=& 2 (\delta C+\delta W) + r\,\pd_r (\delta A+\delta B + \delta C+2\,\delta W) + r^2 \,\ol\Box \delta W \nn\\
 \cR_{\zeta\zeta} &=& \cR_{\xi\xi} \sin^2\xi \\
 \cR_{\theta r} &=& \pd_\theta \pd_r (\delta A+\delta B+2\,\delta W) +\frac2r \, \pd_\theta (\delta C+\delta W) + \cot\theta\,\pd_r \delta B \nn\\
 \cR_{\theta\theta} &=& -1  + 2\, a_0^2\,\ol\Box_2 \delta B +\pd_\theta^2 (\delta A-\delta C+2\,\delta W-\delta B) \nn\\
 \cR_{\varphi\varphi} &=& e^{2\ol B} \bigg[ - 1 - 2\,\delta B +a_0^2\ol\Box \delta B  + \cot\theta \,\pd_\theta (\delta A-\delta C+2\,\delta W+\delta B) \bigg] \,, \nn
\eea
where $\ol\Box = \ol\Box_4 + \ol\Box_2$ simplifies when restricted to functions $f(\theta,r)$, with
\be
 \ol\Box_4 f(\theta,r) = \bigg[ \pd_r^2 + \frac2r \,\pd_r \bigg] f(\theta,r) \quad \hbox{and} \quad
 \ol\Box_2 f(\theta,r) = \bigg[ \frac1{a_0^2}\left(\pd_\theta^2 + \cot\theta \,\pd_\theta\right) \bigg] f(\theta,r) \,.
\ee

This leads to the following set of 8 independent linear field equations, whose solutions we seek. The dilaton and Maxwell equations are
%
%1
\bea \label{app:eq1}
 \hbox{(dilaton:)}&& \;\;
 \Bigl(\pd_\theta^2+\cot\theta \, \pd_\theta\Bigr) \delta\phi +a_0^2\(\pd_r^2+\frac2r\pd_r\)\delta\phi =  \(\delta\phi+\delta B-\frac{\delta F_{\theta\varphi}}{\ol F_{\theta\varphi}} \) \\
%2
\label{app:eq2}
 \hbox{(Maxwell:)}&& \;\;
 \frac{a_0^2}{r^2}\pd_r\(r^2\frac{\delta F_{r\varphi}}{\ol F_{\theta\varphi}}\)+\pd_\theta\(\frac{\delta F_{\theta\varphi}}{\ol F_{\theta\varphi}}-\delta B-\delta\phi +\delta A-\delta C+2\,\delta W\)=0 \,,
\eea
while the trace-reversed Einstein equations linearize to
%3
\bea
\label{app:eq3}
 \hbox{($r\theta$)}&&
 \frac{\delta F_{r\varphi}}{\ol F_{\theta\varphi}}=-\pd_\theta\pd_r\(\delta A+\delta B+2\,\delta W\)-\frac2r\pd_\theta(\delta C+\delta W) -\cot\theta \, \pd_r\delta B \\
%4
\label{app:eq4a}
 \hbox{($tt$)}&&
 \(\pd_\theta^2+\cot\theta \, \pd_\theta + a_0^2\pd_r^2+\frac{2a_0^2}r\pd_r\)\delta A= \frac12 \(\delta\phi+\delta B-\frac{\delta F_{\theta\varphi}}{\ol F_{\theta\varphi}} \) \\
%5
\label{app:eq5}
 \hbox{($rr$)}&&
 -\Bigl(\pd_\theta^2+\cot\theta \, \pd_\theta\Bigr)\delta C+\frac{2a_0^2}r\pd_r(\delta C+2\,\delta W)+a_0^2\pd_r^2\(\delta A+\delta B+2\,\delta W\)\nn\\
 &&\qquad\qquad\qquad\qquad\qquad\qquad\qquad\qquad\qquad\qquad =-\frac1{2}\(\delta\phi+\delta B-\frac{\delta F_{\theta\varphi}}{\ol F_{\theta\varphi}}\) \\
%6
\label{app:eq6}
 \hbox{($\xi\xi$)}&&
 \frac{2a_0^2}{r^2}(\delta C+\delta W)+\frac{a_0^2}r\pd_r\(\delta A+\delta B+\delta C+2\,\delta W\) +a_0^2\ol\square \delta W \nn\\
 &&\qquad\qquad\qquad\qquad\qquad\qquad\qquad\qquad\qquad\qquad =-\frac1{2}\(\delta\phi+\delta B-\frac{\delta F_{\theta\varphi}}{\ol F_{\theta\varphi}}\) \\
%7
 \label{app:eq7}
 \hbox{($\theta\theta$)}&&
 \Bigl(\pd_\theta^2+2\cot\theta \, \pd_\theta\Bigr)\delta B + \pd_\theta^2\Bigl(\delta A-\delta C+2\delta W\Bigr)=  \frac1{2}\(\delta\phi+3\delta B-3\frac{\delta F_{\theta\varphi}}{\ol F_{\theta\varphi}}\) \\
%8
 \label{app:eq8}
 \hbox{($\varphi\varphi$)}&&
 \Bigl(\pd_\theta^2+2\cot\theta \, \pd_\theta\Bigr)\delta B +a_0^2\(\pd_r^2+\frac2r\pd_r\)\delta B +\cot\theta \, \pd_\theta\Bigl(\delta A-\delta C+2\delta W\Bigr) \nn\\
 && \qquad\qquad\qquad\qquad\qquad\qquad\qquad\qquad\qquad\qquad =\frac1{2}\(\delta\phi+3\delta B-3\frac{\delta F_{\theta\varphi}}{\ol F_{\theta\varphi}}\) .
\eea
In what follows it is also useful to combine eqs.~\pref{app:eq1} and \pref{app:eq4a} to obtain
%4
\be
\label{app:eq4}
 \hbox{($tt$ and dilaton)} \quad
 \(\pd_\theta^2+\cot\theta \, \pd_\theta + a_0^2\pd_r^2+\frac{2a_0^2}r\pd_r\)\(2\,\delta A+\delta\phi\)=0 \,.
\ee

\subsection*{Far-field Solutions}

We now turn to finding the general solutions to these equations, with our interest primarily in the far-field regime corresponding to distances much further than the KK scale: $r \gg a_0$. This region is well within the  4D regime, and the dominant result should therefore fall off like $1/r$ and so we expand all of the perturbations in a series in $1/r$,
\ba \label{appeq:rexpansion}
 \delta\phi = H_0(\theta)+\frac{H_1(\theta)}r + \cdots\,, &&\quad
 \delta B =  B_0(\theta)+\frac{ B_1(\theta)}r + \cdots\,, \nn\\
  \delta A = A_0(\theta)+\frac{A_1(\theta)}r + \cdots\,, &&\quad
 \delta C = C_0(\theta)+\frac{C_1(\theta)}r + \cdots\\
 \delta W = W_0(\theta)+\frac{W_1(\theta)}r + \cdots\,, &&\quad
 \delta F_{\ssM\ssN} = \delta F^{(0)}_{\ssM\ssN}(\theta) + \frac1r \, \delta F^{(1)}_{\ssM\ssN}(\theta) + \cdots\,, \nn
\ea
and solve the equations neglecting terms of order $1/r^2$ and higher.

\medskip\noindent{\em Eliminating the gauge field}

\medskip\noindent
Equations \pref{app:eq2} and \pref{app:eq3} above can be combined to eliminate the gauge field strength in terms of the metric and dilaton perturbations. To this end, notice that differentiating \pref{app:eq3} gives
\ba \label{appeq:Frth}
 \frac{a_0^2}{r^2}\pd_r\(r^2\frac{\delta F_{r\varphi}}{\ol F_{\theta\varphi}}\)&=&-\frac{a_0^2}{r^2}\pd_r\(r^2\left[\pd_\theta\pd_r(\delta A+\delta B+2\,\delta W) + \frac2r\pd_\theta (\delta C+\delta W)+\cot\theta \, \pd_r\delta B\right]\)\nn\\
 &=&-\frac{a_0^2}{r^2}\pd_r\Bigl(-\pd_\theta(A_1+ B_1+2\,W_1)+2r\pd_\theta (C_0+W_0)+2\pd_\theta (C_1+W_1) - \cot\theta \,  B_1\Bigr)\nn\\
 &=&-\frac{2a_0^2}{r^2}\pd_\theta (C_0+W_0) \,,
\ea
which can be neglected in what follows because it is order $1/r^2$. Using this in equation \pref{app:eq2} then yields
\be
 \pd_\theta\(\frac{\delta F_{\theta\varphi}}{\ol F_{\theta\varphi}}-\delta B-\delta\phi +\delta A-\delta C+2\delta W \)=0 \,,
\ee
which may be integrated to give
\be
 \frac{\delta F_{\theta\varphi}}{\ol F_{\theta\varphi}}=\delta B+\delta\phi-\delta A+\delta C - 2\delta W +\widetilde Q \,.
\ee
Here we introduce the notation $\widetilde Q (r) = \widetilde Q_0 + \widetilde Q_1/r + \cdots$ where $\widetilde Q_i$ are $\theta$-independent integration constants.

\medskip\noindent{\em Determining $\delta A- \delta C+2 \,\delta W$}

\medskip\noindent
Next, subtracting equations \pref{app:eq7} and \pref{app:eq8}, and using $\ol\Box_4 \delta B = 0$ (which follows from eqs.~\pref{appeq:rexpansion}), gives
\be
 \pd_\theta^2\(\delta A-\delta C+2\delta W\)=\cot\theta \, \pd_\theta\(\delta A-\delta C+2\delta W\) \,,
\ee
which integrates to give
\be \label{appeq:ACWeq}
 \delta A - \delta C + 2 \delta W = \tilde A + \hat A \cos\theta \,,
\ee
where, as before, $\tilde A(r) = \tilde A_0 + \tilde A_1/r + \cdots$, $\hat A (r) = \hat A_0 + \hat A_1/r + \cdots$, where $\tilde A_i$ and $\hat A_i$ are integration constants.

\medskip\noindent{\em The dilaton equation}

\medskip\noindent
Using the $1/r$ expansion and the result for the Maxwell field we find eq.~\pref{app:eq1} implies the dilaton perturbation satisfies
\ba
 \Bigl( \pd_\theta^2+\cot\theta \, \pd_\theta\Bigr) \(\delta\phi_0+\frac{\delta\phi_1}r\) &=& \delta A - \delta C + 2 \, \delta W - \widetilde Q \nn\\
 &=&\tilde A -\widetilde Q + \hat A_0\cos\theta \,.
\ea
This result can be integrated to obtain
\be
 \delta\phi = \tilde H + \hat H \ln\(\frac{1-\cos\theta}{\sin\theta}\) +(\widetilde Q -\tilde A) \ln \sin\theta  - \frac12 \hat A \cos\theta \,,
\ee
with integration constants $\tilde H_i$ and $\hat H_i$, grouped into the $r$-dependent combinations $\tilde H(r) = \tilde H_0 + \tilde H_1/r + \cdots$ and $\hat H(r) = \hat H_0 + \hat H_1/r + \cdots$.

\medskip\noindent{\em Solving for $\delta B$}

\medskip\noindent\label{solveB}
Next we use equation \pref{app:eq7}, eliminating as before the gauge field on the right hand side:
\bea
 \Bigl(\pd_\theta^2+2\cot\theta \, \pd_\theta\Bigr)\delta B &=& -\pd_\theta^2(\delta A-\delta C + 2\delta W)+ \left[ -\delta\phi+\frac32(\delta A-\delta C+2\delta W)-\frac32 \, \widetilde Q \right] \,\nn\\
 &=& -\tilde H -\hat H \ln\(\frac{1-\cos\theta}{\sin\theta}\)+ (\tilde A - \widetilde Q) \(\frac32+\ln \sin\theta \) +3\hat A \cos\theta  \,.\nn\\
\eea
Following \cite{LargeDim} we write the integral
\ba
 \delta B &=& \widetilde B + \hat B \cot\theta + \frac{\tilde H} 2 \, \theta\cot\theta - \hat H \cM_2(\theta) + (\tilde A - \widetilde Q) \left[ \cH_2(\theta)-\frac34 \, \theta\cot\theta \right] -\hat A \cos\theta \,,\nn\\
\ea
with integration constants $\widetilde B(r) = \widetilde B_0 + \widetilde B_1/r + \cdots$, $\hat B(r) = \hat B_0 + \hat B_1/r + \cdots$, and the functions $\cH_2$ and $\cM_2$ defined by:
\ba
 \cM_2(x) := \int_0^x \exd y \; \frac{\cM_1(y)}{\sin^2 y} \quad
 &&\hbox{with} \quad  \cM_1(x) := \int_0^x\exd y \; \sin^2y\,\ln\(\frac{1-\cos y }{\sin y}\) \nn\\
 \cH_2(x) := \int_0^x\exd y \; \frac{\cH_1(y)}{\sin^2 y } \quad &&\hbox{with} \quad  \cH_1(x) := \int_0^x\exd y \; \sin^2 y\, (\ln \sin y)  \,.
\ea
Notice that because $\ln[(1-\cos \theta)/\sin \theta] = \ln[\tan(\theta/2)]$ changes sign under $\theta \to \pi - \theta$, it follows that $\cM_1(\pi - x) = \cM_1(x)$ and so for small $x$
\be
 \cM_1(x) = \cM_1(\pi - x) \simeq \frac{x^3}{3} \left[ \ln \left( \frac{x}{2} \right) - \frac13 \right] \,.
\ee
As a result $\cM_2(x)$ is well-behaved at both $x = 0$ and $x = \pi$, and $\cM_2(\pi - x) = \cM_2(\pi) - \cM_2(x)$ with
\be
 \cM_2(x) \simeq \frac{x^2}{6} \left[ \ln \left( \frac{x}{2} \right) - \frac16 \right] \,,
\ee
for small $x$.

\medskip\noindent{\em Disentangling $\delta W$ and $\delta C$.}

\medskip\noindent
We next use eqn.~\pref{app:eq6}. The important thing to notice here is that on the left-hand side the only terms that are not order $1/r^2$ come from $\ol\Box_2 \delta W$. Dropping the other terms and eliminating the gauge field as above we get
\bea
 \Bigl(\pd_\theta^2+\cot\theta \, \pd_\theta\Bigr)\delta W &=& -\frac12\(\delta A-\delta C+2\delta W\)+\frac12 \, \widetilde Q\nn\\
 &=&\frac12 \(\widetilde Q - \tilde A \) - \frac12 \hat A \cos\theta \,.
\eea
The general solution is
\be
 \delta W = \widetilde W + \widehat W \ln\(\frac{1-\cos\theta}{\sin\theta}\) +\frac12 \Bigl(\tilde A - \widetilde Q\Bigr) \ln \sin\theta + \frac14\hat A\cos\theta \,,
\ee
with 4 new integration constants $\widetilde W_i$ and $\widehat W_i$, combined as before into the $r$-dependent combinations $\widetilde W(r) = \widetilde W_0 + \widetilde W_1/r + \cdots$ and $\widehat W(r) = \widehat W_0 + \widehat W_1 /r + \cdots$.

An identical argument holds for eq.~\pref{app:eq5}, which we solve for $\delta C$ leading to the solution
\be
 \delta C = \tilde C + \hat C \ln\(\frac{1-\cos\theta}{\sin\theta}\) -\frac12\Bigl( \tilde A-\widetilde Q \Bigr) \ln \sin\theta - \frac14\hat A\cos\theta \,,
\ee
with integration constants $\tilde C(r) = \tilde C_0 + \tilde C_1/r + \cdots$ and $\hat C(r) = \hat C_0 + \hat C_1/r + \cdots$. Combined with eq.~\pref{appeq:ACWeq} these expressions for $\delta C$ and $\delta W$ also determine $\delta A$ through:
\be
 \delta A = \delta C - 2\delta W + \tilde A + \hat A \cos\theta \,.
\ee

\medskip\noindent{\em Evaluating the gauge fields}

\medskip\noindent
The expressions used earlier to eliminate the gauge fields can now be used to evaluate them explicitly. First, we see from eq.~\pref{app:eq3} that
\ba
 \frac{\delta F_{r\varphi}}{\ol F_{\theta\varphi}}&=&-\frac{2}r\pd_\theta (\delta C + \delta W) + O(1/r^2)=-\frac{2(\hat C_0+\hat W_0)}{r \sin\theta} \,.
\ea
The expression for the $\delta F_{\theta\phi}$ similarly becomes:
\ba
 \frac{\delta F_{\theta\varphi}}{\ol F_{\theta\varphi}}&=&\delta B+\delta\phi-\delta A+\delta C-2\delta W+\widetilde Q \nn\\
 &=& \widetilde B + \hat B \cot\theta + \tilde H \( 1+\frac12\theta\cot\theta\) + \hat H \(\ln \( \frac{1-\cos\theta}{\sin\theta}\) - \cM_2(\theta) \) \nn\\
 &&+(\widetilde Q-\tilde A)\(1+\ln(\sin\theta) + \frac34 \, \theta \cot\theta- \cH_2(\theta)\) -\frac52\hat A \cos\theta\,.
\ea

Finally, the only equation not yet used is eq.~\pref{app:eq4}, and inserting the above expressions for $\delta A$ and $\delta \phi$ into this provides the additional condition
\be
 \tilde A(r) = \widetilde Q(r) \,.
\ee

\medskip\noindent{\em Coordinate conditions}

\medskip\noindent
We are free to set a few of the remaining integration constants to specific values using some residual coordinate freedom. Following \cite{LargeDim} we use the freedom to shift $\theta$ to put the position of one brane at $\theta = \theta_+ = 0$. Since the brane locations correspond to positions where the metric coefficient $e^B = \alpha \sin \theta \, (1 + \delta B)$ vanishes, this is achieved by asking
\be
  \hat B(r) = \hat B_0 + \frac{\hat B_1}{r} = 0 \,.
\ee
In principle we wish to do so for all $r$, but because $\hat B(r)$ depends on $r$ setting $\hat B_1 = 0$ requires shifting $\theta$ by an $r$-dependent amount: $\theta \to \theta - \theta_0 - \theta_1/r + \cdots$. Although this shift in general introduces off-diagonal $\exd \theta \exd r$ terms into the metric (and so takes us outside of the {\em ansatz} used to solve the field equations), the cross terms arise at order $1/r^2$ and so can be neglected in the far-field regime.

\medskip\noindent{\em Solutions summary}

\medskip\noindent
We now collect the final expressions for the general perturbations:
\be
 \delta \phi = \tilde H(r) + \hat H(r) \ln\(\frac{1-\cos\theta}{\sin\theta}\) -\frac12\hat A (r) \cos\theta \,,
\ee
\be
 \delta A = \tilde A(r) + \tilde C(r) -2 \widetilde W(r) + [ \hat C(r) -2 \,\widehat W(r) ] \ln\(\frac{1-\cos\theta}{\sin\theta}\) + \frac14 \hat A(r) \cos\theta \,,
\ee
\be
 \delta C = \tilde C(r) + \hat C(r) \ln\(\frac{1-\cos\theta}{\sin\theta}\) -\frac14\hat A(r) \cos\theta \,,
\ee
\be
 \delta B = \widetilde B(r) + \frac{ \tilde H(r)}{2} \, \theta \cot\theta - \hat H(r) \, \cM_2(\theta) - \hat A(r) \cos\theta\,,
\ee
\be
 \delta W_0= \widetilde W(r) + \widehat W(r) \, \ln\(\frac{1-\cos\theta}{\sin\theta} \) +\frac14\hat A(r) \, \cos\theta\,,
\ee
\be
 \frac{\delta F^{(0)}_{\rho\varphi}}{\ol F_{\rho\varphi}} = \widetilde B(r) + \tilde H(r) \(1+\frac\theta 2 \, \cot\theta\) + \hat H(r) \left[ \ln\(\frac{1-\cos\theta}{\sin\theta}\) -\cM_2(\theta) \right] - \frac52 \hat A(r) \, \cos\theta \,,
\ee
\be
 \hbox{and} \qquad
 \frac{\delta F_{r\varphi}}{\ol F_{\rho\varphi}} = - \frac{2}{r} \left( \frac{\hat C_0+\hat W_0}{\sin\theta} \right) \,.
\ee

\subsection*{Boundary conditions}

We next specify the values of the various integration constants by imposing the appropriate boundary conditions.

\medskip\noindent{\em Near-brane matching conditions}

\medskip\noindent
We first fix the integration constants using the near-brane boundary conditions, which in the far-field region are given by eqs.~\pref{branematch1} and \pref{branematch2}.

Of these, eqs.~\pref{branematch1} requires the near-brane limit, $\lim_{\theta \to 0} \theta \partial_\theta$, to vanish for the functions $\delta A$, $\delta C$, $\delta W$ and $\delta \phi$. Furthermore, these conditions hold for all $r$. They therefore require the following integration constants to vanish:
\be
 \hat H(r) = \hat C(r) = \widehat W(r) = 0 \,.
\ee
These imply the solutions reduce to the simpler form
\be
 \delta \phi = \tilde H(r) -\frac12\hat A (r) \cos\theta \,,
\ee
\be
 \delta A = \tilde A(r) + \tilde C(r) - 2 \widetilde W(r) + \frac14 \hat A(r) \cos\theta \,,
\ee
\be
 \delta C = \tilde C(r)  -\frac14\hat A(r) \cos\theta \,,
\ee
\be \label{appeq:dBwbc}
 \delta B = \widetilde B(r)  + \frac{ \tilde H(r)}{2} \, \theta \cot\theta - \hat A(r) \cos\theta\,,
\ee
\be
 \delta W_0=\widetilde W(r) +\frac14\hat A(r) \, \cos\theta\,,
\ee
\bea
 \frac{\delta F_{\rho\varphi}}{\ol F_{\rho\varphi}} &=& \widetilde B(r) + \tilde H(r) \(1+\frac\theta 2 \, \cot\theta\)  - \frac52 \hat A(r) \, \cos\theta \nn\\
 &=& \delta B + \delta \phi - \hat A(r) \cos \theta \,,
\eea
and $\delta F_{r\varphi} = 0$.

Next, eq.~\pref{branematch2} states (for all large $r$)
\be \label{branematch2app}
 \lim_{\theta \to 0} \Bigl(\partial_\theta e^B - 1 \Bigr) = -\frac{\kappa^2 L_+}{2\pi} \quad \hbox{and} \quad
 \lim_{\theta \to \theta_-} \Bigl(- \partial_\theta e^B - 1 \Bigr) = -\frac{\kappa^2 L_-}{2\pi} \,,
\ee
where, because the initial rugby ball solution requires identical branes, we choose $L_+ = L_- := L$. These conditions relate integration constants in $\delta B$ to the properties of the source branes.

The three quantities, $\widetilde B$, $\tilde H$ and $\hat A$ appearing in $\delta B$ at this point have a simple physical interpretation in terms of the geometrical properties of the extra dimensions, which we briefly outline to better understand the implications of eqs.~\pref{branematch2app}. These three quantities can be traded for an ($r$-dependent) change to the defect angles at each brane, and a change in the proper distance between the two branes. In particular the positions of the branes are set by the two places, $\theta = \theta_\pm$, where $e^{B(\theta_\pm)} \simeq \alpha \sin \theta (1 + \delta B)$ vanishes, which for the perturbed solution found above occurs at $\theta_+ = 0$ (because of the coordinate choice made earlier) and $\theta_- = \pi + \delta \theta$. Similarly, the conical defect angles at each brane are given by the derivatives of $e^B$ at these positions:
\be
 \alpha_\pm = \pm \left. \partial_\theta e^B \right|_{\theta_\pm}
 = \pm \alpha \left[ \left(1 + \widetilde B + \frac12 \, \tilde H\right) \cos \theta_\pm - \frac12 \, \tilde H \, \theta_\pm \sin \theta_\pm - \hat A \cos 2\theta_\pm \right] \,.
\ee
Using these explicit expressions we can solve to obtain
\be
 \theta_- \simeq \pi \left( 1 - \frac12 \, \tilde H \right) \, \quad
 \hbox{and} \quad
 \frac{\delta\alpha_\pm}{\alpha} = \widetilde B + \frac12 \, \tilde H \mp \hat A \,.
\ee
Consequently $\tilde H$ captures the change, $\delta \ell$, in the proper distance between the branes,
\be
 \frac{\delta \ell }{ \ell} = \frac{\delta \theta }{\pi} = \frac12 \, \tilde H \,,
\ee
while $\widetilde B$ describes a common change to the defect angle at both branes, and $\hat A$ gives a relative change to these two defect angles.

\end{document}